\documentclass[journal]{IEEEtran}
\IEEEoverridecommandlockouts
\usepackage{cite}
\usepackage{amsmath,amssymb,amsfonts}
\usepackage{graphicx}
\usepackage{textcomp}
\usepackage{xcolor}
\usepackage{siunitx}
\usepackage[T1]{fontenc}
\usepackage{subfigure}
\usepackage{comment}

\def\BibTeX{{\rm B\kern-.05em{\sc i\kern-.025em b}\kern-.08em
    T\kern-.1667em\lower.7ex\hbox{E}\kern-.125emX}}

\def\Tpk{T_{{\sf p}}[k]}
\def\tck{t_{{\sf c}}[k]}
\def\tsk{t_{{\sf s}}[k]}
\def\tek{t_{{\sf e}}[k]}
\def\bias{\bar{b}}

\newcounter{mytempeqncnt}

\begin{document}

\title{Comprehensive Signal Modeling for Talkative Power Conversion}

\author{\mbox{ }
\thanks{\mbox{ }}%
\thanks{\mbox{ }}}

\author{Jan Mietzner, {\it Senior Member, IEEE},~ Cerikh Chakraborty, {\it Graduate Student Member, IEEE}, \\ Peter A.~Hoeher, {\it Fellow, IEEE}, and Lutz Lampe, {\it Senior Member, IEEE}
\thanks{Manuscript version v1.0 (April 15, 2025).}%
\thanks{J. Mietzner is with the Faculty of Design, Media and Information, Research \& Transfer Center Digital Reality, Hamburg University of Applied Sciences (HAW), D-20099 Hamburg, Germany (e-mail:
jan.mietzner@haw-hamburg.de) and with Kiel University, Faculty of Engineering, Chair of Information and Coding Theory, D-24143 Kiel, Germany.

C. Chakraborty and L. Lampe are with the Department of Electrical \& Computer Engineering (ECE), The University of British Columbia (UBC), Vancouver, BC V6T 1Z4, Canada (e-mail: cc2211@ece.ubc.ca, lampe@ece.ubc.ca).

P. A. Hoeher is with Kiel University, Faculty of Engineering, Chair of Information and Coding Theory, D-24143 Kiel, Germany (e-mail: ph@tf.uni-kiel.de).}}

\maketitle

\begin{abstract} 
Talkative power conversion is a switching ripple communication technique that integrates data modulation into a switched-mode power electronics converter, enabling simultaneous information transmission and power conversion. 
Despite numerous research papers published over the last decade on various theoretical and practical aspects of this emerging topic, thorough signal modeling suitable for analysis and computer simulations is still lacking. 
In this article, we derive the continuous-time output voltage of a DC/DC switched-mode power electronics converter for a broad range of pulsed-based modulation schemes. We also develop corresponding discrete-time signal models and assess their accuracies. Finally, we devise a generic end-to-end signal model for arbitrary modulation signals, discuss implications of continuous-time and discrete-time signal modeling on equalization, and consider generalizations to include parasitic effects as well as the influence of general impedance loads. 
\end{abstract}

\begin{IEEEkeywords}
Channel modeling, talkative power conversion, power line communication, pulse width modulation converters, simultaneous information and power transfer. 
\end{IEEEkeywords}


\section{Introduction} 
Developing the future energy grid, known as the smart grid, is one of the most significant challenges facing our society  \cite{dileep2020survey}. With the steadily increasing share of renewable energy sources, power electronic converters (PECs) have become core components of the smart grid \cite{bose2017power}. PECs bridge energy sources, electrical storage systems, and electrical consumers, necessitating enhanced  communication. Therefore, embedded communication is a key function of the smart grid  \cite{dileep2020survey,abrahamsen2021communication}, enabling applications such as decentralized power control and telemetry, and many others.

Communication in energy grids can be established either through external units or by using existing lines. The former includes wireless communication methods such as wireless mesh networks, cellular radio, optical fiber communication links, and digital subscriber lines \cite{abrahamsen2021communication}. The latter  involves technologies like power line communication (PLC) \cite{Cano2016,Lampe2016} and power-over-Ethernet (PoE) \cite{Maniktala2013}. PLC enables wide-area networking over transmission and distribution grids often using narrowband data communication, as well as local-area networking with broadband data communication. PoE is a reuse technique for joint power supply and communication networking within indoor environments.

This paper addresses an alternative communication technique called talkative power conversion (TPC). TPC is a ripple communication technique that integrates data modulation  into the power conversion process, enabling simultaneous information transmission and power conversion. This approach embeds communication directly into switched-mode PECs, naturally combining power electronics and digital communication.

This innovative technology originated in \cite{Stefanutti2008} as a ``switching frequency modulation'' technique and was extended in \cite{Wu2015} as a ``power/signal dual modulation'' scheme. The names ``power talk'' and ``talkative power'' have been coined in \cite{Popovski2016} (focusing on system control) and \cite{He2020} (focusing on the power converter), respectively. ``Talkative power conversion'' is a more generalized terminology that considers all open systems interconnection (OSI) communication layers \cite{Liserre2023}.  The various designations also include ``zero-additional hardware PLC'' \cite{Han2022}. Despite the appeal of this name, additional hardware is typically still necessary, because the data need to be recovered at the receiver side.  More importantly, TPC is not restricted to PLC applications like data transmission over direct current (DC) or alternating current (AC) power grids and electric vehicle (EV) charging. It also offers opportunities for wireless applications such as wireless power charging and visible-light communication (VLC) using light-emitting diodes (LEDs) \cite{He2020, Liserre2023, Wang2023}.

The number of publications in the field of TPC has increased significantly in recent years. Many contributions have focused on sophisticated PEC topologies \cite{Du2017, Choi2017, Han2022}, on various modulation schemes \cite{Wu2015, Wang2017, Hoeher2021, Chen2022, Zhang2022b, Leng2022, Chen2023, Moench2023}, on coding \cite{Chen2021, Hoeher2021} and  multiuser system aspects \cite{Popovski2016, Popovski2017}, on improving the networking capabilities \cite{Wu2019, Zhang2022a}, and on voltage control \cite{Zhu2019, Wu2023}. 
TPC has been suggested for a wide range of applications \cite{Rodriguez2018, Loose2018, Zhang2019BMS, Zhu2019, Qian2019, Rodriguez2020, Yu2020, Hua2021, Rodriguez2023, Wu2023}. 
Beyond embedding renewable energy sources into the electric grid and eliminating conventional transformers, PECs are used in residential, industrial, vehicular, and avionic microgrids, for EV and wireless charging, and in various consumer products, to name just a few use cases. Further, more recent publications on TPC may be found in \cite{Liu2024a}--\cite{Leng2025dc}. 

Interestingly, despite the relatively large number of contributions made in the area of TPC so far, the fundamental scientific question of signal modeling for joint information and power transfer based on TPC has never been solved comprehensively. To the best of our knowledge, despite many years of research on TPC, only \cite{Chen2021} and \cite{Hoeher2021} have presented signal models so far. Correspondingly, the primary objective of this paper is to provide comprehensive signal models for TPC, so as to close known gaps in the literature and to devise a solid basis for corresponding receiver designs and upcoming system optimizations. Our main contributions are summarized in the following subsection. 

\subsection{Paper Scope, Organization, and Main Contributions}
In this paper, we devise a comprehensive signal model concerning the information-carrying output voltage of a switched-mode PEC under the TPC paradigm. 
For concreteness and practical relevance, we focus on the example of a step-down DC/DC voltage converter with an $LC$ lowpass filter and an Ohmic load. 
The analysis is, however, applicable to any DC/DC converter where the switching unit and the filter unit can be decoupled. 
With additional effort it can be generalized to other DC/DC converters as well.
The proposed signal modeling can be applied to any switching pattern and thus to any TPC-based power/signal modulation method.

The main contributions of this paper are as follows:
\begin{itemize}
\item Starting from a basic circuit model of a synchronous 
step-down DC/DC converter, we consider the underlying differential equation (DE) and derive the continuous-time output voltage for a broad range of pulsed-based modulation schemes. 
Numerical simulation results illustrate and verify our analytical findings. This is the first time that a comprehensive continuous-time signal model for TPC is presented. Although continuous-time signaling was considered in \cite{Chen2021}, comprehensive equations regarding  signal modeling aspects and the influence of the underling circuit and modulation parameters, including the influence of initial conditions, were omitted.  
\item We systematically establish corresponding discrete-time signal models, which place a known iterative update model from \cite{Hoeher2021} in a wider context. 
In particular, significant insights are gained by assessing the accuracy of these models as a function of the sampling rate. This is the first time, that the accuracy of discrete-time signal models for TPC is assessed in a rigorous fashion. Although discrete-time signal modeling was considered in \cite{Hoeher2021}, the accuracy of the proposed model could not be assessed, due to the lack of a comprehensive continuous-time signal model.
\item We extend the discrete-time models to an asynchronous step-down DC/DC converter, which is subject to a noncontinuous mode of operation. This is the first time that signal modeling for TPC using asynchronous step-down DC/DC converters is addressed.
\item We devise a generic end-to-end signal model, which is valid for -- in principle -- arbitrary input signals, including effects due to non-ideal switching units (e.g., causing finite signal slopes). Based on the proposed continuous-time and discrete-time signal models, we provide a conceptional discussion, how our findings can be utilized for developing corresponding receiver designs, especially regarding suitable filter- or state-based equalization schemes. This is the first time that systematic guidelines are devised regarding receiver designs for future TPC applications. These guideline also provide a starting point for corresponding performance analyses.
\item  We generalize the TPC topology to include parasitic effects and arbitrary impedance loads and derive the corresponding continuous-time signal models. This is the first time that comprehensive continuous-time signal models are devised for such general settings.
\end{itemize}
The signal modeling under consideration provides a common framework for the fields of power electronics and communications engineering.
The proposed {\em signal modeling} can be used as a basis for {\em system modeling}, i.e., for TPC system design, for circuit design, for corresponding evaluations, and for optimizations. 
This includes, but is not limited to 
\begin{itemize}
\item the design and selection of modulation schemes (including duty-cycle and data-dependent switching pattern optimizations, investigation of ripple characteristics and the impact of the ripple on performance measures like the total harmonic distortion and power quality), 
\item advanced power converter topologies (including multi-layer topologies, bidirectional communication, and higher-order filters), 
\item application-dependent channel impairments and receiver design (including equalization, synchronization and channel estimation, quantization, bit error rate analysis, signal-to-noise ratio calculation, transient behavior and intersymbol interference, noise immunity, jitter, impact of load variations, latency), 
\item networking aspects (including multiple access technologies and network topologies), and  
\item system comparisons (including TPC vs.~PLC system benchmarking). 
\end{itemize}
System aspects are beyond the scope of this paper, however.

The outline of the paper is as follows: The considered PEC topology is introduced in Section~\ref{II}. Continuous-time signal modeling for TPC is addressed in Section~\ref{III}, whereas corresponding discrete-time signal models are devised in Section~\ref{IV}. The generic end-to-end signal model is discussed Section~\ref{V}. Implications of the proposed continuous-time and discrete-time signal modeling on equalization are addressed in Section~\ref{VI-new}. Generalizations to parasitic effects and arbitrary impedance loads are presented in Section~\ref{VII-new}. 
Finally, the paper is concluded in Section~\ref{VI}, discussing possible extensions and interesting directions for future work.     

Our modeling approach covers the two parameters that are most important from the point of view of both power electronics and communication engineering: output voltage and inductor current. These parameters have a strong impact on power quality and current stress. Both parameters are captured continuously over time, hence peak and average values are inherently included. Thus, the modeling is suitable for power electronics circuit design as well as for communication-related analysis and simulations.
The continuous-time signal model in Section~\ref{III} is useful for studying the overall behavior regarding the data communication functionality and its interaction with power electronics (e.g., for different settings of the circuit parameters). In particular, important implications for corresponding transmitter and receiver designs can be gained from this model, as pointed out in Section~\ref{VI-new}.
On the other hand, the discrete-time signal models presented in Section~\ref{IV} may serve as a basis for performance evaluation of the data communication part using (Monte Carlo) computer simulations. 
Finally, the generic end-to-end signal model presented in Section~\ref{V} may serve as a basis for developing corresponding equalization schemes, in order to mitigate intersymbol interference (ISI) effects.

Regarding the existing literature, we note that while \cite{Chen2021}  considered the same step-down DC/DC converter type as in this paper, the authors did not provide a comprehensive signal model, neither continuous-time nor discrete-time. In \cite{Hoeher2021}, the focus was on corresponding discrete-time signal modeling, suitable receiver processing, and an experimental proof of concept, but an in-depth characterization of the continuous-time (end-to-end) signal model was out of scope. Our paper therefore aims to complement the available literature, by devising a comprehensive signal model for step-down DC/DC voltage converters, while generalizations to other power converter topologies are possible.
Our signal model is relevant for a wide range of applications. 
This includes, for example,  data transmission over power grids with Ohmic loads as well as VLC, where the output voltage of a PEC drives an LED, which converts the corresponding driver current into a (proportional) light signal \cite[Ch.~3]{Hoeher2019}.

\subsection{Experimental Verification}
An experimental verification of the signal models discussed within the scope of this paper was done in \cite{Hoeher2021} based on the iterative update model proposed therein. Therefore, we refrain from including any (repetitive) experimental results. Rather, the objective of this paper is to extend and deepen theoretical results regarding TPC signal modeling.

\subsection{Mathematical Definitions}
For better readability, we identify  continuous-time signals using parentheses and time variable $t\in\mathbb{R}$, e.g. $x(t)$, and brackets for  discrete-time signals and time index $k\in\mathbb{Z}$ or $n\in\mathbb{Z}$, e.g. $x[n]$.
 $\delta_0(t)$ denotes a Dirac impulse at \mbox{$t=0$}.
 ${\rm rect}(t)$ denotes an ideal rectangular pulse function with unit amplitude and width, centered around \mbox{$t=0$}.
 Correspondingly, ${\rm rect}((t-t_{\sf c})/T_{\sf p})$ is an ideal rectangular pulse with unit amplitude and width $T_{\sf p}$, centered around $t=t_{\sf c}$. $\Theta(t)$ denotes the Heaviside function, which is given by $\Theta(t)=0$ for $t<0$ and $\Theta(t)=1$ for $t\geq 0$. Vectors are represented by bold-face symbols, e.g. $\mathbf{x}$. We use $\dot{x}(t)$, $\ddot{x}(t)$, and $\dddot{x}(t)$ as short-hand notations for the derivatives ${\rm d}x(t)/{\rm d}t$, ${\rm d}^2x(t)/{\rm d}t^2$, and ${\rm d}^3x(t)/{\rm d}t^3$, respectively. Finally, $a\ast b$ denotes the linear convolution of signals $a$ and~$b$, $[x]_+ = \max(x,0)$ denotes the maximum of $x$ and zero, and ${\rm j}=\sqrt{-1}$ is the imaginary unit.

\section{Topology Under Investigation}\label{II} 

\begin{figure}[t]
\centering
\includegraphics[width=0.5\textwidth]{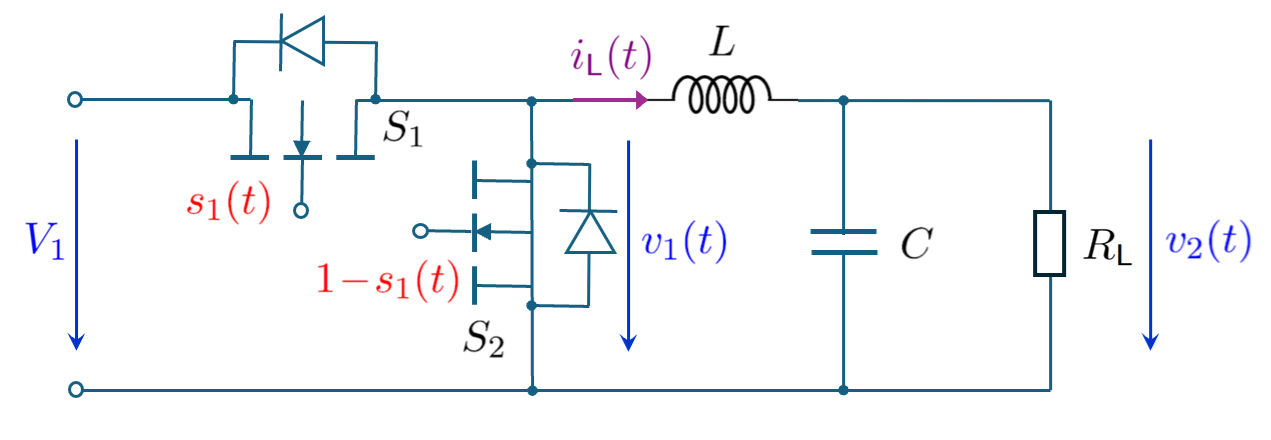}
\caption{Circuit diagram of the considered synchronous DC-DC step-down converter with coupled switches, $LC$ lowpass filter, and Ohmic load (cf.~\cite{Hoeher2021}).} 
\label{fig:circuit_buck}
\end{figure}

Fig.~\ref{fig:circuit_buck} displays the circuit diagram of a synchronous DC-DC step-down converter (also referred to as {\it buck converter}), which converts an (average) input voltage level $V_1$ into a smaller (average) output voltage level $V_2<V_1$ \cite{Hoeher2021}. 
For the ease of exposition and without loss of generality, we assume a fixed input voltage $V_1$. 
The voltage conversion is done by means of two coupled electronic switches $S_1$ and $S_2$, e.g.\ metal-oxide-semiconductor field-effect transistors (MOSFETs) or Gallium Nitride (GaN) semiconductors.
The switches $S_1$ and $S_2$ are controlled by a two-level signal $s_1(t)$ with duty cycle \mbox{$\delta<1$}. 
Traditionally, this pulse width modulation (PWM) signal is periodic. 
During an active pulse, switch $S_1$ is closed, and switch $S_2$ is open. 
Vice versa, when $s_1(t)$ is equal to zero, switch $S_1$ is open, and switch $S_2$ is closed. 
In the following, the switches $S_1$ and $S_2$ are assumed to be ideal. Consequently, the signal $s_1(t)$ can be modeled as an ideal square-wave signal given by~\cite[Ch.~4.3]{Hoeher2019}
\begin{equation}\label{eq:s1}
    s_1(t) = \sum_{k=-\infty}^{+\infty} {\rm rect}\left( \frac{t-kT-t_{\sf c}}{T_{\sf p}}\right),
\end{equation}
where $T:=1/f_{\sf s}$ denotes the period of $s_1(t)$, $f_{\sf s}$ the switching frequency, $T_{\sf p}:=\delta \cdot T$ the pulse duration, and $t_{\sf c}$ the center of the pulse within the $k$th period, which is given by $t_{\sf c} = T/2$ in the unmodulated case. 
Based on the above definitions, the  voltage delivered by the switching unit can be expressed as $v_1(t):= V_1\cdot s_1(t)$.
Subsequently, the DC-DC step-down converter employs an analog lowpass filter, which is often  a second-order lowpass composed of an inductor  and a capacitor, as shown in Fig.~\ref{fig:circuit_buck}. 
Correspondingly, the output voltage $v_2(t)$ is a smoothed version of the voltage $v_1(t)$ at the input of the filter. 
Throughout, we assume that an Ohmic load $R_{\sf L}$ is connected to the output port of the DC-DC step-down converter. 
By design, for a synchronous buck converter the inductor current $i_{\sf L}(t)$ is always flowing, hence the converter is said to operate in continuous conduction mode (CCM). 
This holds for arbitrary values of $R_{\sf L}$. 

The DC-DC step-down converter can be made ``talkative'' by embedding an information sequence into the switching signal $s_1(t)$.
However, care must be taken not to compromise the primary function of voltage conversion, by keeping fluctuations of the output voltage $v_2(t)$ -- denoted as {\it ripple}~-- to a minimum, in order to preserve power quality. 
The ripple voltage $v_{2,{\sf ripple}}(t)=v_2(t)-V_2=v_2(t)-\delta\cdot V_1$ is a deterministic function of $s_1(t)$ and is hence information-carrying as well.
A simple method, which requires no or little additional hardware components, is to modify the rectangular pulses within the switching signal $s_1(t)$ regarding their frequency, widths and/or respective positions \cite{Hoeher2021}. 
To this end, the same microcontroller is sufficient, which controls the switches $S_1$ and $S_2$.  
Examples include variable pulse-width modulation (VPWM), where the pulse width $T_{\sf p}$ is varied as a function of digital data symbols $x[k]$ ($\Tpk:=f(x[k])$) -- subject to an (adjustable) average duty cycle of $\delta$ -- and variable pulse-position modulation (VPPM), where the pulse center is varied accordingly ($\tck:=g(x[k])$). 
Correspondingly, the symbol duration is given by the period $T$ of the unmodulated switching signal. 
Regarding the added communication function, the step-down converter in Fig.~\ref{fig:circuit_buck} simultaneously adopts the role of the transmitter, which converts the discrete-time data symbols $x[k]$ into an analog output voltage $v_2(t)$. 
Accordingly, $v_2(t)$ (or more precisely $v_{2,{\sf ripple}}(t)$) represents the transmitted signal and will be the focus of our modeling approach. 

The received signal can be written as 
\begin{equation}
r(t) = c\cdot v_2(t) + n(t),
\end{equation}
where $c$ denotes a proportionality factor and $n(t)$ an additive noise term. 
Subsequent analog-to-digital conversion (ADC) will then enable digital signal processing steps for recovering the transmitted data symbols $x[k]$ \cite{Hoeher2021}.
For example, in the case of a power grid setting, the received signal $r(t)$ corresponds to the voltage measured at the Ohmic load, which is identical to $v_2(t)$ (cf.~Fig.~\ref{fig:circuit_buck}) or a scaled version thereof. 
In a VLC setting, the output voltage $v_2(t)$ is associated with a corresponding driver current $i_{\sf F}(t)=v_2(t)/R_{\sf L}$, which is converted by the LED into an optical signal. 
In this case, the Ohmic load $R_{\sf L}$ comprises the series resistor and the operation-point dependent intrinsic resistance of the LED. 
At the receiver, a photodiode converts the impinging light signal into a proportional photo current $i_{\sf P}(t)$. 
Correspondingly, in this case the received signal $r(t)$ corresponds to $i_{\sf P}(t)$, and $c$ is a proportionality factor associated with electrical-to-optical and optical-to-electrical conversion  \cite[Ch.~3]{Hoeher2019}.

\section{Continuous-Time Signal Modeling for TPC}\label{III}
In this section, we derive a continuous-time (i.e., analog) signal model concerning the output voltage $v_2(t)$ of the step-down converter for the assumption of an embedded pulse-based modulation scheme (e.g., VPWM or VPPM).  
To this end, we analyze the output voltage $v_2(t)$ for given initial conditions and arbitrary data symbols $x[k]$ and cast the resulting expression for $v_2(t)$ in terms of a basic pulse shape $g_{\sf tx}(t)$, which comprises both the modulation characteristics and the circuit parameters of the step-down converter. 
Corresponding discrete-time approximations of the analog signal model will be considered in Section~\ref{IV}.

Basic analysis of the step-down converter circuit in Fig.~\ref{fig:circuit_buck} yields the following equations for the derivatives of the inductor current $i_{\sf L}(t)$ and the output voltage $v_2(t)$:
\begin{equation}\label{eq:dot_iL}
    \frac{{\rm d}i_{\sf L}(t)}{{\rm d}t} = \frac{1}{L}\bigg( v_1(t) - v_2(t) \bigg),
\end{equation}
\begin{equation}\label{eq:dot_v2}
    \frac{{\rm d}v_2(t)}{{\rm d}t} = \frac{1}{C}\bigg( i_{\sf L}(t) - \frac{v_2(t)}{R_{\sf L}} \bigg).
\end{equation}
Regarding the relationship between the filter input and output voltages, the following DE is obtained:
\begin{equation}
    v_1(t) = v_2(t) + \frac{L}{R_{\sf L}}\cdot \frac{{\rm d}v_2(t)}{{\rm d}t} + LC\cdot \frac{{\rm d}^2v_2(t)}{{\rm d}t^2}.
\end{equation}
For arbitrary data symbols $x[k]$ embedded in $v_1(t)$, the DE can be solved by transforming it to the Laplace domain, according to
\begin{eqnarray}\label{eq:V1_s}
    V_1(s) &=& V_2(s) + \frac{L}{R_{\sf L}} \bigg(s\cdot V_2(s)-v_2(0)\bigg) \nonumber \\ 
           && + LC \bigg(s^2\cdot V_2(s)-s\cdot v_2(0) - \dot{v}_2(0) \bigg). 
\end{eqnarray}
Note that (\ref{eq:V1_s}) contains the initial conditions $v_2(0)$ and $\dot{v}_2(0)$ for the output voltage and its derivative at $t=0$. 
Via (\ref{eq:dot_v2}), these can be translated to a corresponding initial condition for the inductor current, according to $i_{\sf L}(0)= v_2(0)/R_{\sf 
 L} + C\cdot \dot{v}_2(0)$.
Resolving (\ref{eq:V1_s}) for $V_2(s)$ yields
\begin{equation} \label{eq:Laplace_V2}   
    V_2(s)=\frac{V_1(s)+L\big(Cs+1/R_{\sf L}\big)\cdot v_2(0) + LC\cdot \dot{v}_2(0)}{LC\cdot \big( s-s_{0,1}\big)\big(s-s_{0,2}\big)},
\end{equation}
where
\begin{equation}
    s_{0,1/2}=-\frac{1}{2CR_{\sf L}} \pm \sqrt{\left(\frac{1}{2CR_{\sf L}}\right)^2 - \frac{1}{LC}}.
\end{equation}
In the following, we assume that the Ohmic load meets the condition $R_{\sf L}>0.5 \sqrt{L/C}$. 
In this case, the roots $s_{0,1/2}$ are distinct and complex-conjugated poles of $V_2(s)$. 

Subsequently, let $\tsk=h(x[k])$ denote the data-dependent start of the pulse within the $k$th symbol period and $\tek$ the corresponding end of the pulse. 
For example, in the case of VPWM, we have $\tsk=T/2-\Tpk/2$ and $\tek=T/2+\Tpk/2$. 
Similarly, for VPPM we have $\tsk=\tck-T_{\sf p}/2$ and $\tek=\tck+T_{\sf p}/2$.
With these definitions, and constraining $s_1(t)$ to a right-sided signal over $K$ symbol periods, $V_1(s)$ is obtained in closed form as
\begin{equation}\label{eq:V1_s_expl}  
    V_1(s)= \frac{V_1}{s}\sum_{k=0}^{K-1} \left({\rm e}^{-s\tsk} - {\rm e}^{-s\tek}\right)\cdot {\rm e}^{-skT}.
\end{equation}
The details regarding the derivation of (\ref{eq:V1_s_expl}) as well as the subsequent derivations are relegated to the Appendix. Based on (\ref{eq:V1_s_expl}), the corresponding closed-form expression for $V_2(s)$ is given by (\ref{eq:V2_s_expl}).
\begin{figure*}[t!] 
\normalsize
\setcounter{mytempeqncnt}{\value{equation}}
\setcounter{equation}{9}
\begin{equation}\label{eq:V2_s_expl}
    V_2(s) \;=\; \frac{(s+1/(R_{\sf L}C))\cdot v_2(0) + \dot{v}_2(0)}{(s-s_{0,1})(s-s_{0,2})} \;+\; \frac{V_1}{LC\cdot s(s-s_{0,1})(s-s_{0,2})}\sum_{k=0}^{K-1} \left({\rm e}^{-s\tsk} - {\rm e}^{-s\tek}  \right)\cdot {\rm e}^{-skT}
\end{equation}
\setcounter{equation}{10}
\hrulefill
\vspace*{4pt}
\end{figure*}
The first term in (\ref{eq:V2_s_expl}) concerns the transient behavior of the output voltage $v_2(t)$ starting from the initial conditions $v_2(0)$ and $\dot{v}_2(0)$. 
It can be rewritten in terms of a partial fraction expansion with denominator terms $(s-s_{0,1})$ and $(s-s_{0,2})$, and the resulting sum terms can then be transformed back to time domain by using known Laplace transform pairs. 
The transient signal component of $v_2(t)$ thus follows as
\begin{eqnarray}\label{eq:v2t_t}
    v_{2,{\sf t}}(t)&=&{\rm e}^{-at}\bigg( \frac{ a\cdot v_2(0) + \dot{v}_2(0)}{b}\cdot {\rm sin}(bt)   \nonumber \\
    & &  \;\;\;\;\;\;\;\;\;\;\;\;+ v_2(0)\cdot {\rm cos}(bt)\bigg)\cdot \Theta(t),
\end{eqnarray}
where $a:=-{\rm Re}\{s_{0,1/2}\}$ and $b:={\rm Im}\{s_{0,1}\}=-{\rm Im}\{s_{0,2}\}$. 
Note that $v_{2,{\sf t}}(t)$ vanishes, if both $v_2(0)$ and $\dot{v}_2(0)$ are equal to zero. 
The second term in (\ref{eq:V2_s_expl}) represents the data-dependent behavior of the output voltage $v_2(t)$. 
It can be transformed back into the time domain using known Laplace transform pairs. Furthermore, noting that the resulting convolution of ${\rm e}^{s_{0,1}t}\cdot \Theta(t)$ with ${\rm e}^{s_{0,2}t}\cdot \Theta(t)$ in time domain can be expressed as $({\rm e}^{s_{0,1}t}-{\rm e}^{s_{0,2}t})/(s_{0,1}-s_{0,2}) \cdot \Theta(t)$, the remaining convolution integral can be solved in closed form. The data-dependent signal component of $v_2(t)$ thus results as
\begin{eqnarray}\label{eq:v2d_t}
    v_{2,{\sf d}}(t)&=&V_1\cdot \left( 1-{\rm e}^{-at} \cdot \left({\rm cos}(bt) + \frac{a}{b}\cdot {\rm sin}(bt) \right)\right)\cdot \Theta(t) \nonumber \\
    && \hspace*{-0.5cm}\ast \left( \sum_{k=0}^{K-1}  \delta_0(t-kT-\tsk) - \delta_0(t-kT-\tek)\right).\nonumber\\
\end{eqnarray}
By exchanging the convolution operation and the summation in (\ref{eq:v2d_t}), the expression for $v_{2,{\sf d}}(t)$ may be represented as
\begin{equation}\label{eq:v2d_pulses}
    v_{2,{\sf d}}(t) = V_1 \cdot \sum_{k=0}^{K-1} g_{\sf tx}(t-kT-t_{\sf c}[k])
\end{equation}
with basic pulse shape $g_{\sf tx}(t)$ as given in (\ref{eq:gtx_t}).
\begin{figure}[t]
\centering
\includegraphics[width=0.5\textwidth]{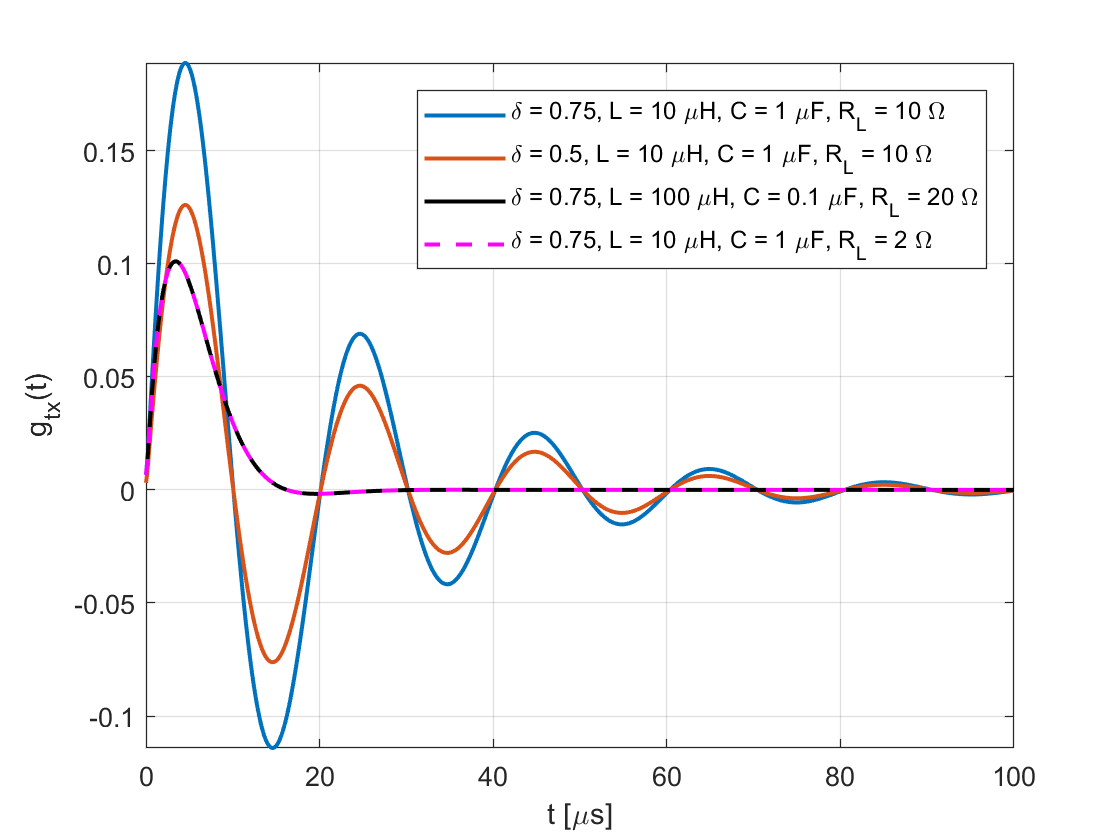}
\caption{Illustration of the basic pulse shape $g_{\sf tx}(t)$ for the unmodulated case ($t_{\sf c} = T_{\sf p}/2$, $T_{\sf p}=\delta \cdot T$), given the parameter values in Table~\ref{tab:num-para}.} 
\label{fig:gtx}
\end{figure}
\begin{figure*}[t!] 
\normalsize
\setcounter{mytempeqncnt}{\value{equation}}
\setcounter{equation}{13}
\begin{eqnarray}\label{eq:gtx_t}
 g_{\sf tx}(t)&=&\left( 1-{\rm e}^{-a(t+T_{\sf p}[k]/2)} \cdot \left({\rm cos}(b(t+T_{\sf p}[k]/2)) + \frac{a}{b}\cdot {\rm sin}(b(t+T_{\sf p}[k]/2) \right)\right)\cdot \Theta(t+T_{\sf p}[k]/2)\nonumber \\
 &&-\left( 1-{\rm e}^{-a(t-T_{\sf p}[k]/2)} \cdot \left({\rm cos}(b(t-T_{\sf p}[k]/2)) + \frac{a}{b}\cdot {\rm sin}(b(t-T_{\sf p}[k]/2) \right)\right)\cdot \Theta(t-T_{\sf p}[k]/2)
\end{eqnarray}
\setcounter{equation}{14}
\hrulefill
\vspace*{4pt}
\end{figure*}
Note that (\ref{eq:v2d_pulses})/(\ref{eq:gtx_t}) comprise both the modulation characteristics -- in terms of data-dependent variables $t_{\sf c}[k]$ (e.g., for VPPM) and $T_{\sf p}[k]$ (e.g., for VPWM) -- and the circuit parameters of the step-down converter circuit ($L,C,R_{\sf L}$). In particular, (\ref{eq:gtx_t}) may thus serve as a basis to assess the overall amount of ISI due to the modulation scheme {\it and} the circuit parameters.   
The final closed-form expression for the output voltage $v_2(t)$ is given by 
\begin{equation}\label{eq:v2_final}
    v_2(t) = v_{2,{\sf t}}(t) + V_1 \cdot \sum_{k=0}^{K-1} g_{\sf tx}(t-kT-t_{\sf c}[k]).
\end{equation}
In the following, our findings are illustrated and validated by means of selected numerical examples.

\begin{table}[!t]
    \centering
    \caption{Parameter values employed for numerical examples.}
    \begin{tabular}{|c|p{3.7cm}|c|}\hline
     {\bf Parameter} & {\bf Description} & {\bf Value} \\ \hline
     $L$    & Inductance & 10, 100~$\mu$H  \\ \hline
     $C$    & Capacitance & 0.1, 1~$\mu$F \\ \hline
     $R_{\sf L}$    & Ohmic load & 2, 10, 20~$\Omega$ \\ \hline 
     $f_{\sf s}$    & Switching frequency & 1~MHz \\ \hline
     $T$    & Symbol duration  & 1~$\mu$s \\ \hline    
     $\delta$    & Duty cycle  & 0.5, 0.75 \\ \hline
     $V_1$  & Input voltage & arbitrary \\ \hline
     \end{tabular}
    \label{tab:num-para}\vspace*{-3ex}
\end{table}

\begin{figure}[t]
\centering
\includegraphics[width=0.5\textwidth]{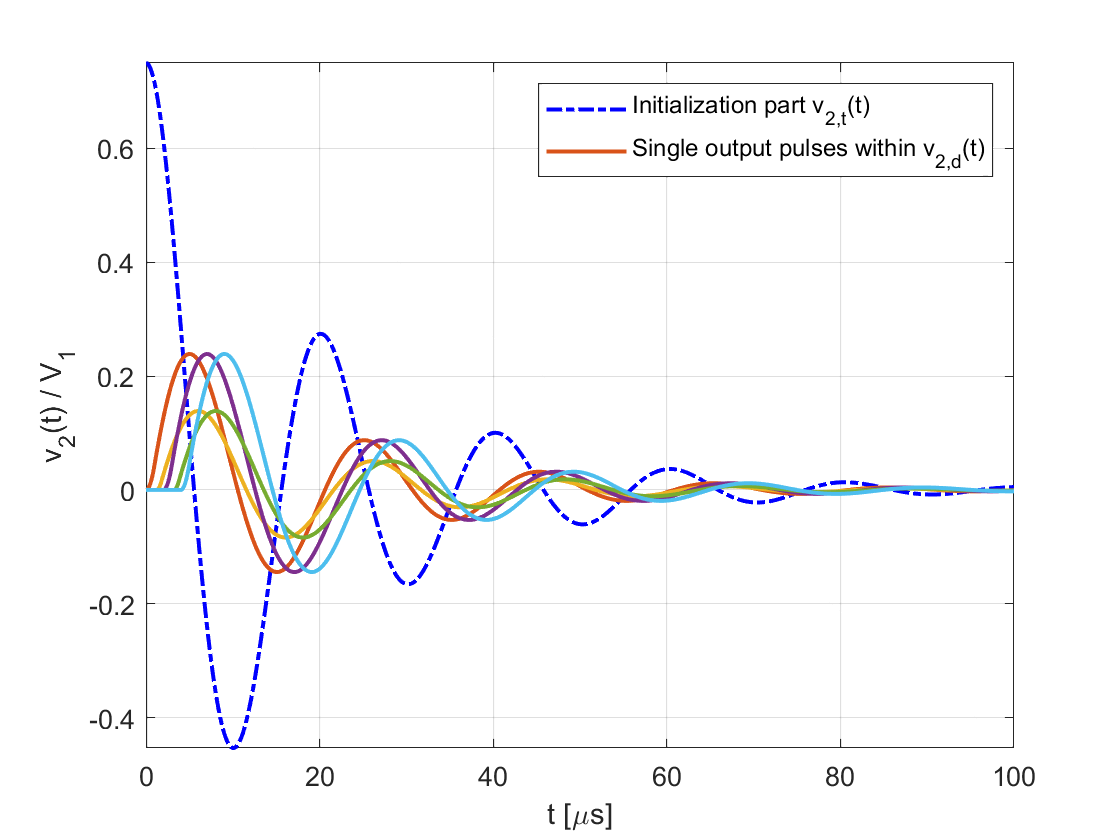}
\caption{Illustration of the transient component (dashed line, blue color) and the data-dependent components (solid lines, mixed colors) of the output voltage $v_2(t)$ for the parameter values in Table~\ref{tab:num-para} (\mbox{$L\!=\!10$~$\mu$H}, \mbox{$C\!=\!1$~$\mu$F}, \mbox{$R_{\sf L}\!=\!10$~$\Omega$}), an arbitrary fixed input voltage $V_1$, initial conditions $v_2(0)/V_1=\delta=0.75$, $\dot{v}_2(0)=0$~V/s, and VPWM for an example data sequence $x[k]=[1,0,1,0,1]$. The duty cycle variation depth was chosen as $\pm 0.2$ for illustrative purposes.} 
\label{fig:v2_expl}
\end{figure}

\begin{figure}[t]
\centering
\includegraphics[width=0.5\textwidth]{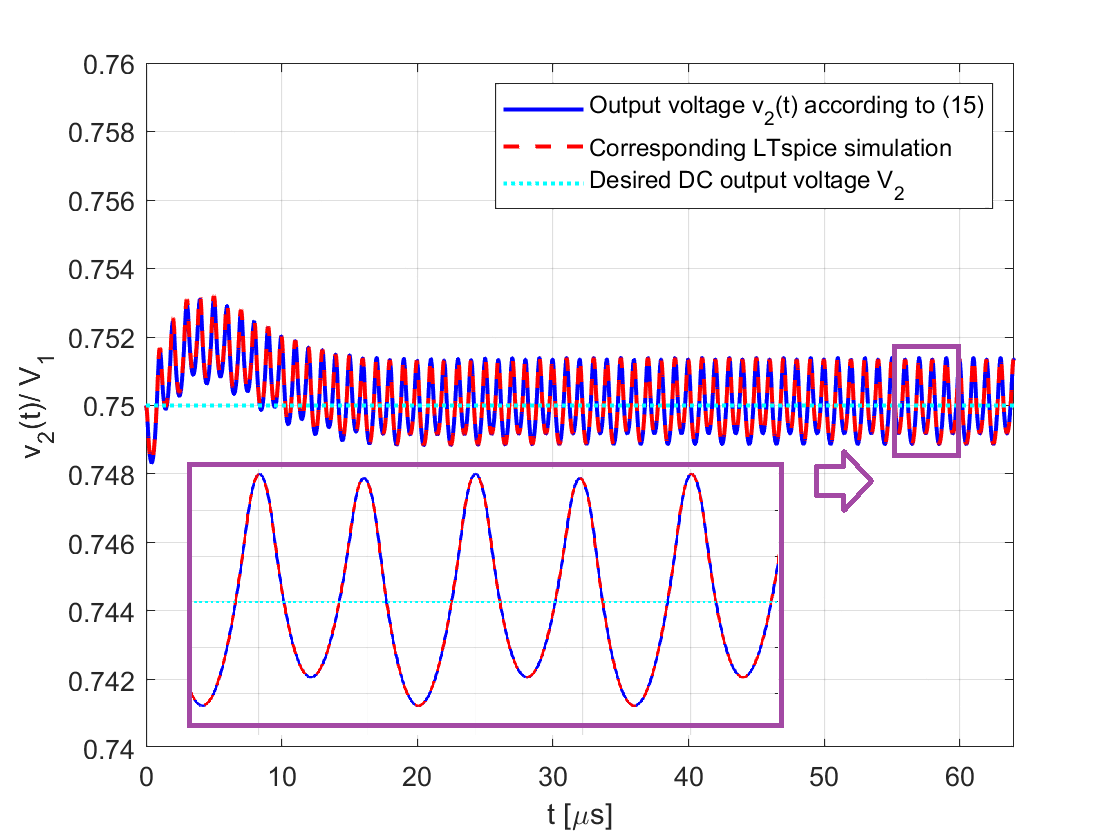}
\caption{Illustration of the overall output voltage $v_2(t)$ according to (\ref{eq:v2_final}) (solid line, blue color) compared against the result of a corresponding LTspice\textsuperscript{\textregistered} simulation (dotted line, red color) for the parameter values in Table~\ref{tab:num-para}  (\mbox{$L\!=\!100$~$\mu$H}, \mbox{$C\!=\!0.1$~$\mu$F}, \mbox{$R_{\sf L}\!=\!20$~$\Omega$}), an arbitrary fixed input voltage $V_1$, initial conditions $v_2(0)/V_1=\delta=0.75$, $\dot{v}_2(0)=0$~V/s, and VPWM for an example data sequence $x[k]=[1,0,1,0,1,...]$ of length 64. The duty cycle variation depth was chosen as $\pm 0.025$.} 
\label{fig:v2_total_LTspice}
\end{figure}

\subsection{Numerical Examples}
As an example, the basic pulse shape $g_{\sf tx}(t)$ resulting for the unmodulated case ($t_{\sf c} = T_{\sf p}/2$, $T_{\sf p}=\delta \cdot T$) and the values listed in Table~\ref{tab:num-para} is illustrated in Fig.~\ref{fig:gtx}. As can be seen, the resulting shape of $g_{\sf tx}(t)$ depends significantly on the choice of the modulation parameters (here, the duty cycle $\delta$) {\it and} the employed circuit parameters ($L,C,R_{\sf L}$). An illustration of the transient and data-dependent components of the output voltage $v_2(t)$ for the example of an arbitrary fixed input voltage $V_1$, \mbox{$L\!=\!10$~$\mu$H}, \mbox{$C\!=\!1$~$\mu$F}, \mbox{$R_{\sf L}\!=\!10$~$\Omega$}, initial conditions $v_2(0)/V_1=\delta=0.75$, $\dot{v}_2(0)=0$~V/s, and VPWM for an example data sequence $x[k]=[1,0,1,0,1]$ is shown in Fig.~\ref{fig:v2_expl}, displaying the first five transmit pulses $g_{\sf tx}(t)$. For illustrative purposes, the variation depth of the duty cycle was chosen as $\pm 0.2$, i.e., for a data bit 0 a duty cycle of 0.55 was employed and for a data bit 1 a duty cycle of 0.95. Typically, practicable values of the duty cycle variation depth are smaller, in order to limit the resulting ripple component of the output voltage $v_2(t)$. Regarding the primary task of power conversion, the magnitude of the ripple component is usually expected to be within the range of a few percent of the nominal output voltage $V_2$. 

Fig.~\ref{fig:v2_total_LTspice} illustrates the overall output voltage $v_2(t)$ according to (\ref{eq:v2_final}) (solid line, blue color) compared against the result of a corresponding LTspice\textsuperscript{\textregistered} simulation (dashed line, red color) for the parameter values in Table~\ref{tab:num-para} (\mbox{$L\!=\!100$~$\mu$H, \mbox{$C\!=\!0.1$~$\mu$F}, \mbox{$R_{\sf L}\!=\!20$~$\Omega$})}, an arbitrary fixed input voltage $V_1$, initial conditions $v_2(0)/V_1=\delta=0.75$, $\dot{v}_2(0)=0$~V/s, and VPWM for an example data sequence $x[k]=[1,0,1,0,1,...]$ of length 64. The duty cycle variation depth was chosen as $\pm 0.025$ in this example. As can be seen, the numerical results based on (\ref{eq:v2_final}) and the results from the LTspice\textsuperscript{\textregistered} simulation overlap, which verifies our analysis presented above. Furthermore, it can be seen that, after the transient component has tapered out, the output voltage fluctuates around the desired DC output voltage $V_2$ (as expected), where the relative magnitude of the ripple component is less than 0.2~\% in this example.

\subsection{Ripple Analysis}
\begin{figure}[t]
\centering
\includegraphics[width=0.5\textwidth]{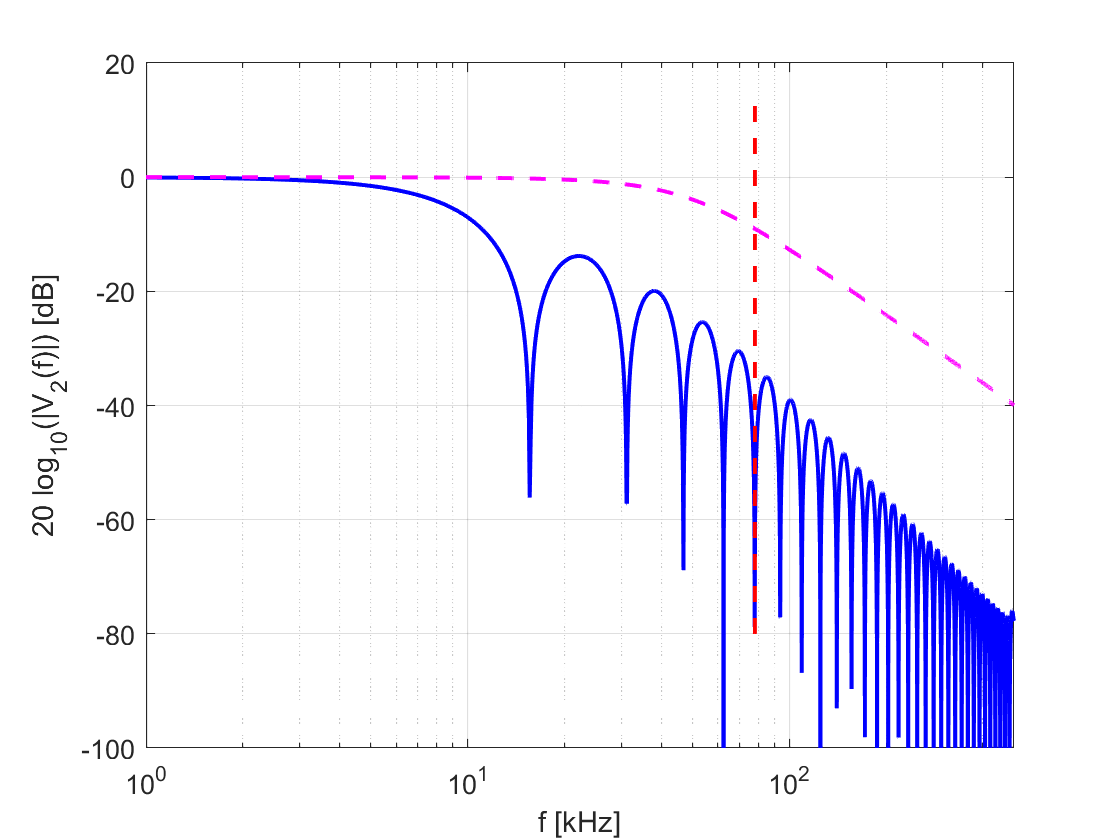}
\caption{Normalized ripple power spectrum $20\log_{10}(|V_{2,{\sf ripple}}(f)|)$  for the parameter values in Table~\ref{tab:num-para} (\mbox{$L\!=\!100$~$\mu$H}, \mbox{$C\!=\!0.1$~$\mu$F}, \mbox{$R_{\sf L}\!=\!20$~$\Omega$}) and VPWM for an example data sequence $x[k]=[1,0,1,0,1,...]$ of length 64 (blue color). The duty cycle was chosen as $\delta=0.75$ and the variation depth as $\pm 0.025$. The cut-off frequency $f_{\sf 3dB}$ of the second-order $LC$ lowpass filter is indicated by the vertical red line. The theoretical frequency response of the second-order $LC$ lowpass filter with Ohmic load $R_{\sf L}$ is included as a reference (magenta color).} 
\label{fig:V2_f_ripple}
\end{figure}

As a by-product of our analysis, the ripple behavior of the output voltage $v_2(t)$ may be analyzed in closed form. To this end, we consider the Laplace transform of the ripple component $v_{2,{\sf ripple}}(t)= v_{2,{\sf d}}(t) - \delta \cdot V_1$, which is given by the first term in (\ref{eq:Laplace_V2}), corrected for the constant term $\delta \cdot V_1$. This yields the following Fourier spectrum: 
\begin{equation}     
    V_{2,{\sf ripple}}(f)=\frac{V_1(f)}{LC\cdot \big( {\rm j}2\pi f-s_{0,1}\big)\big({\rm j}2\pi f-s_{0,2}\big)} - \delta \cdot V_1\cdot\delta_0(f), 
\end{equation}
where
\begin{equation}    
    V_1(f)= \frac{V_1}{{\rm j}2\pi f}\sum_{k=0}^{K-1} \left({\rm e}^{-{\rm j}2\pi f \tsk} - {\rm e}^{-{\rm j}2\pi f \tek}\right)\cdot {\rm e}^{-{\rm j}2\pi f kT}.\\
\end{equation}
As an example, Fig.~\ref{fig:V2_f_ripple} shows the normalized ripple power spectrum $20\log_{10}(|V_{2,{\sf ripple}}(f)|)$ (blue color) for the parameter values in Table~\ref{tab:num-para} (\mbox{$L\!=\!100$~$\mu$H}, \mbox{$C\!=\!0.1$~$\mu$F}, \mbox{$R_{\sf L}\!=\!20$~$\Omega$}, $\delta=0.75$) and VPWM for an example data sequence $x[k]=[1,0,1,0,1,...]$ of length 64. The duty cycle variation depth was chosen as $\pm 0.025$. The cut-off frequency $f_{\sf 3dB}$ of the second-order $LC$ lowpass filter,  
\begin{equation}\label{eq:f_3dB}
    f_{\sf 3dB}=\frac{1}{2\pi}\sqrt{\frac{1+\sqrt{2}}{LC}},
\end{equation} 
has been included as a reference (dashed red line). The low-pass behavior due to the $LC$ filter is clearly visible. The regular shape of the side lobes is due to the chosen, periodic data sequence $x[k]$. However, the observed notches -- especially within the passband of the $LC$ lowpass filter ($f\leq f_{\sf 3dB}$) -- also occur for random data sequences. Furthermore, we have included the theoretical frequency response of the second-order $LC$ lowpass filter with Ohmic load $R_{\sf L}$, which is given by \cite[Ch.~6.2]{Book_LCR-filter}
\begin{equation}\label{eq:Hf}
    H(f)=\frac{1}{1+{\rm j}2\pi f L/ R_{\sf L} - (2\pi f)^2 LC}.
\end{equation}
It displays the known decay of 40~dB per decade for frequencies outside the passband area (magenta color), which is also visible in the ripple power spectrum.

\section{Discrete-Time Signal Modeling for TPC}\label{IV}
The continuous-time signal model derived in Section~\ref{III} is important, in order to understand the behavior of the step-down converter depending on initial conditions and specific TPC modulation parameters. In particular, important implications regarding receiver design considerations can be derived from the continuous-time signal model, as will be discussed in Section~\ref{VI-new}. Corresponding discrete-time signal models, on the other hand, are of interest, because they can readily be implemented on digital computers, in order to simulate  signal characteristics and end-to-end bit-error-rate (BER) performance. 
In the following, we devise different approaches to discrete-time signal modeling and compare them for selected numerical examples. 
Notably, inspiring work in this area was previously presented in \cite{Hoeher2021}. 

\subsection{Discrete-Time Signal Model}
A discrete-time signal model is obtained by applying Euler's method with interval length $\Delta t:= T/J$, where $J$ is referred to as the oversampling factor. Using the notation $x[n]=x(n\Delta t)$, the first-order derivative is approximated by 
\begin{equation}
    \frac{{\rm d}x(t=n\Delta t)}{{\rm d}t} \approx \frac{x[n]-x[n-1]}{\Delta t}.\nonumber
\end{equation} 
The corresponding discrete-time approximations of (\ref{eq:dot_iL}) and (\ref{eq:dot_v2}) follow as 
\begin{equation}\label{eq:dot_iL_discrete}
    i_{\sf L}[n] = i_{\sf L}[n-1] + \frac{\Delta t}{L}\bigg( V_1\cdot s_1[n] - v_2[n] \bigg),
\end{equation}
\begin{equation}\label{eq:dot_v2_discrete}
    v_2[n] = v_2[n-1] + \frac{\Delta t}{C}\bigg( i_{\sf L}[n] - \frac{v_2[n]}{R_{\sf L}} \bigg).
\end{equation} 
Further refinements are possible by using higher-order approximations for the derivative \cite[Ch.~1]{LeVeque2024}. 
In \eqref{eq:dot_iL_discrete},  we have used that $v_1[n]=V_1\cdot s_1[n]$, where $s_1[n]=s_1(n\Delta t)$ is either zero or one, i.e., $s_1[n] \in \{0,1\}$. 
Resolving (\ref{eq:dot_v2_discrete}) for $v_2[n]$ yields
\begin{equation}\label{eq:dot_v2_discrete_acc}
    v_2[n] = \kappa\cdot v_2[n-1] + \mu\cdot i_{\sf L}[n],
\end{equation} 
where 
\begin{equation}
    \kappa:=\frac{CR_{\sf L}}{CR_{\sf L}+\Delta t}, \;\;\; \mu:=\frac{R_{\sf L}\Delta t}{CR_{\sf L}+\Delta t}. 
\end{equation}
Plugging (\ref{eq:dot_v2_discrete_acc}) into (\ref{eq:dot_iL_discrete}) and resolving for $i_{\sf L}[n]$ yields
\begin{equation}\label{eq:dot_iL_discrete_acc}
    i_{\sf L}[n] = \alpha \cdot i_{\sf L}[n-1] + \beta \cdot V_1\cdot s_1[n] + \gamma \cdot v_2[n-1],   
\end{equation}
where
\begin{equation}
    \alpha:=\frac{L(CR_{\sf L}+\Delta t)}{L(CR_{\sf L}+\Delta t)+R_{\sf L}(\Delta t)^2}, \nonumber 
\end{equation}
\begin{equation}
    \beta:=\frac{\Delta t(CR_{\sf L}+\Delta t)}{L(CR_{\sf L}+\Delta t)+R_{\sf L}(\Delta t)^2}, \nonumber
\end{equation}
\begin{equation}
    \gamma:=\frac{-CR_{\sf L}\Delta t}{L(CR_{\sf L}+\Delta t)+R_{\sf L}(\Delta t)^2}.
\end{equation}

\begin{figure}[t]
\centering
\includegraphics[width=0.5\textwidth]{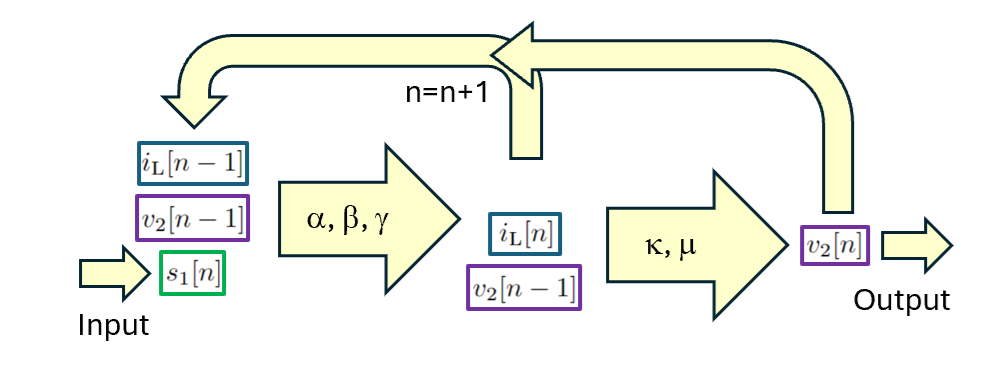}
\caption{Illustration of the iterative approximation (\ref{eq:dot_iL_discrete_acc}), (\ref{eq:dot_v2_discrete_acc}) of the continuous-time inductor current $i_{\sf L}(t)$ and output voltage $v_2(t)$.} 
\label{fig:iteration1}
\end{figure} 

The iterative approximation (\ref{eq:dot_iL_discrete_acc}), (\ref{eq:dot_v2_discrete_acc}) of the continuous-time inductor current $i_{\sf L}(t)$ and output voltage $v_2(t)$ is illustrated in Fig.~\ref{fig:iteration1}. 
Note that within each iteration both previous ($n-1$) {\it and} new ($n$) signal values are required, in order to obtain the new values for $i_{\sf L}[n]$ and $v_2[n]$. 

\subsection{Further Approximations}
The above iteration can be simplified, by revisiting (\ref{eq:dot_iL_discrete}) and (\ref{eq:dot_v2_discrete}) and replacing $v_2[n]$ on the right-hand side by \mbox{$v_2[n-1]$}. 
Note that this constitutes an approximation (\mbox{$v_2[n]\approx v_2[n-1]$}), which is accurate for a sufficiently large oversampling factor $J$ (i.e., a sufficiently small sampling interval $\Delta t$). 
We then obtain
\begin{equation}\label{eq:dot_iL_discrete_approx2021}
    i_{\sf L}[n] = i_{\sf L}[n-1] + \frac{\Delta t}{L} \cdot V_1\cdot s_1[n] - \frac{\Delta t}{L} \cdot v_2[n-1],   
\end{equation}
\begin{equation}\label{eq:dot_v2_discrete_aprox2021}
    v_2[n] = \frac{CR_{\sf L}-\Delta t}{CR_{\sf L}}\cdot v_2[n-1] + \frac{\Delta t}{C}\cdot i_{\sf L}[n],
\end{equation}
i.e., the parameters in the iterative approximation (\ref{eq:dot_iL_discrete_acc}), (\ref{eq:dot_v2_discrete_acc}) reduce to $\alpha=1$, $\beta=\Delta t/L$, $\gamma=-\beta$, $\kappa=(CR_{\sf L}-\Delta t)/(CR_{\sf L})$, and $\mu=\Delta t/C$, while the basic iterative procedure remains unchanged (cf.~Fig.~\ref{fig:iteration1}). 
The iterative approximation (\ref{eq:dot_iL_discrete_approx2021}), (\ref{eq:dot_v2_discrete_aprox2021}) corresponds to the procedure proposed earlier in \cite{Hoeher2021}.

\begin{figure}[t]
\centering
\includegraphics[width=0.35\textwidth]{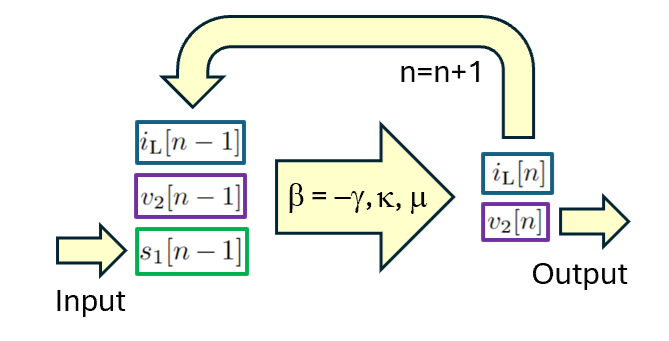}
\caption{Illustration of the iterative approximation (\ref{eq:dot_iL_discrete_approx2024}), (\ref{eq:dot_v2_discrete_aprox2024}) of the continuous-time inductor current $i_{\sf L}(t)$ and output voltage $v_2(t)$.} 
\label{fig:iteration2}
\end{figure}

Conceptually, it is convenient to rewrite (\ref{eq:dot_iL_discrete_approx2021}) and (\ref{eq:dot_v2_discrete_aprox2021}) in the form of a linear prediction model 
\begin{equation}
    x[n] = \xi\cdot x[n-1] + c[n-1],
\end{equation}
where $\xi$ denotes a scaling factor, and $c[n-1]$ is a suitable correction (or update) term, which depends only on previous signal values. 
This can be accomplished by introducing the additional approximations $s_1[n]\approx s_1[n-1]$ in (\ref{eq:dot_iL_discrete_approx2021}) and  $i_{\sf L}[n]\approx i_{\sf L}[n-1]$ in (\ref{eq:dot_v2_discrete_aprox2021}). 
Then, the iterative updates read
\begin{equation}\label{eq:dot_iL_discrete_approx2024}
    i_{\sf L}[n] = i_{\sf L}[n-1] + \beta \cdot V_1\cdot s_1[n-1] + \gamma \cdot v_2[n-1],   
\end{equation}
\begin{equation}\label{eq:dot_v2_discrete_aprox2024}
    v_2[n] = \kappa\cdot v_2[n-1] + \mu\cdot i_{\sf L}[n-1],
\end{equation}
with $\beta, \gamma, \kappa,$ and $\mu$ as specified for (\ref{eq:dot_iL_discrete_approx2021}), (\ref{eq:dot_v2_discrete_aprox2021}).
Note that the approximation $s_1[n]\approx s_1[n-1]$ will merely entail a shift of the output voltage $v_2[n]$ by one sample, since $s_1[n]$ is assumed to be an ideal square-wave signal. 
The approximation $i_{\sf L}[n]\approx i_{\sf L}[n-1]$ is accurate for a sufficiently large oversampling factor $J$.
The simplified iterative approximation based on the linear prediction model (\ref{eq:dot_iL_discrete_approx2024}), (\ref{eq:dot_v2_discrete_aprox2024}) is illustrated in Fig.~\ref{fig:iteration2}.

\subsection{Accuracy Analysis}
In the following, we evaluate the impact of the choice of the interval $\Delta t$ (or the oversampling factor $J$, respectively) on the accuracy of the three discrete-time signal models outlined above. To this end, we evaluate the respective mean squared error (MSE) of the resulting output voltage $v_2[n]$ in comparison to $v_{2,{\sf ref}}[n]:=v_2(n\Delta t)$, where $v_2(t)$ is the solution (\ref{eq:v2_final}) of the DE, after the transient signal component $v_{2,{\sf t}}(t)$ has tapered out. 
Let $N_{\sf s}$ denote the total number of samples evaluated (depending on the choice of $J$).

\begin{figure}[t]
\centering
\includegraphics[width=0.5\textwidth]{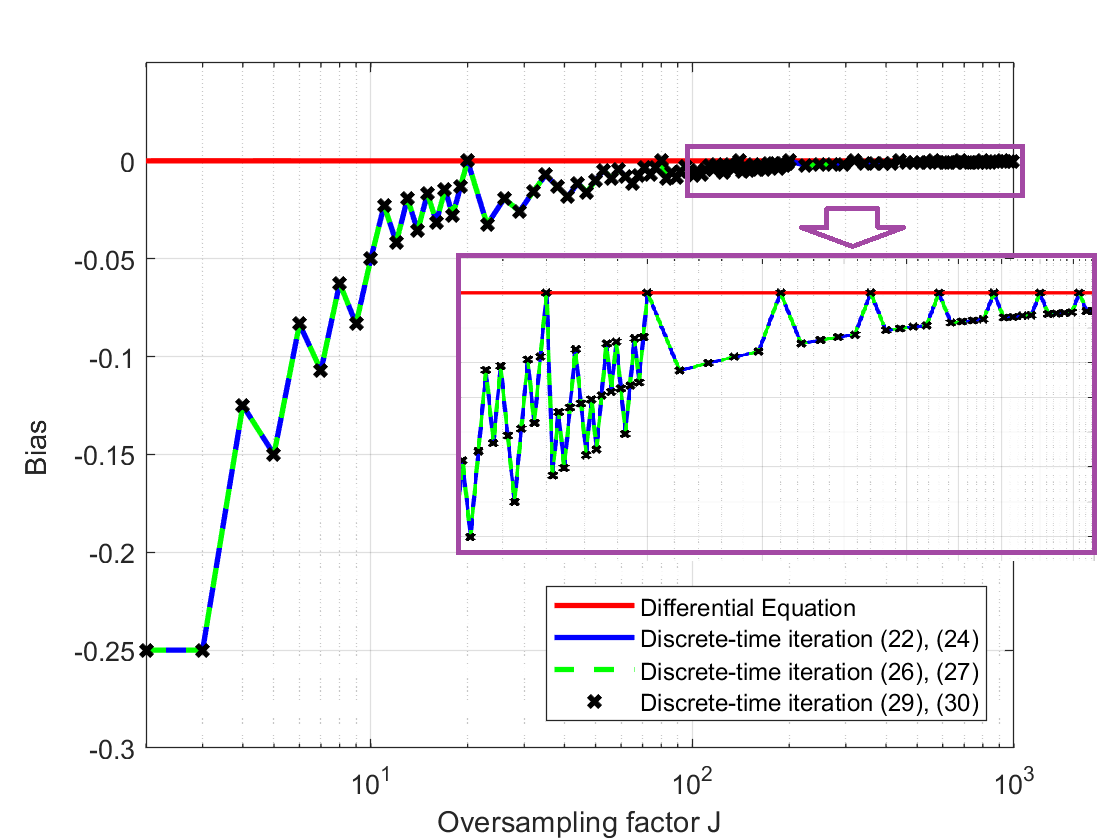}
\caption{Resulting normalized bias $\bias/V_1$ for the three discrete-time approximations as a function of the oversampling factor $J$ for the same setting as in Fig.~\ref{fig:v2_expl} and an example data sequence $x[k]=[1,0,1,0,1,...]$ of length 64.} 
\label{fig:bias}
\end{figure}

A first important observation is that the reference samples $v_{2,{\sf ref}}[n]$ always fluctuate around the mean value equal to the nominal DC output voltage $V_2$, irrespective of the choice of $J$. However, this is not the case for the discrete-time solutions discussed above. In fact, there are particular choices of $J$, for which the corresponding bias 
\begin{equation}
    \bias:= \frac{1}{N_{\sf s}} \sum_{n=0}^{N_{\sf s}-1} v_2[n] - V_2
\end{equation}
is relatively close to zero, whereas other choices of $J$ can entail a notable bias. This can be seen in Fig.~\ref{fig:bias}. While the bias generally tends to zero with increasing oversampling factor $J$ (as expected), there is a significant variation within the bias. In particular, as can be seen in the augmented subfigure, there are favorable ($\bias=0$) and less favorable ($\bias<0$) choices of $J$. Furthermore, we note that the bias $\bias$ is virtually the same for all three discrete-time solutions in this example.  

\begin{figure}[t]
\centering
\includegraphics[width=0.5\textwidth]{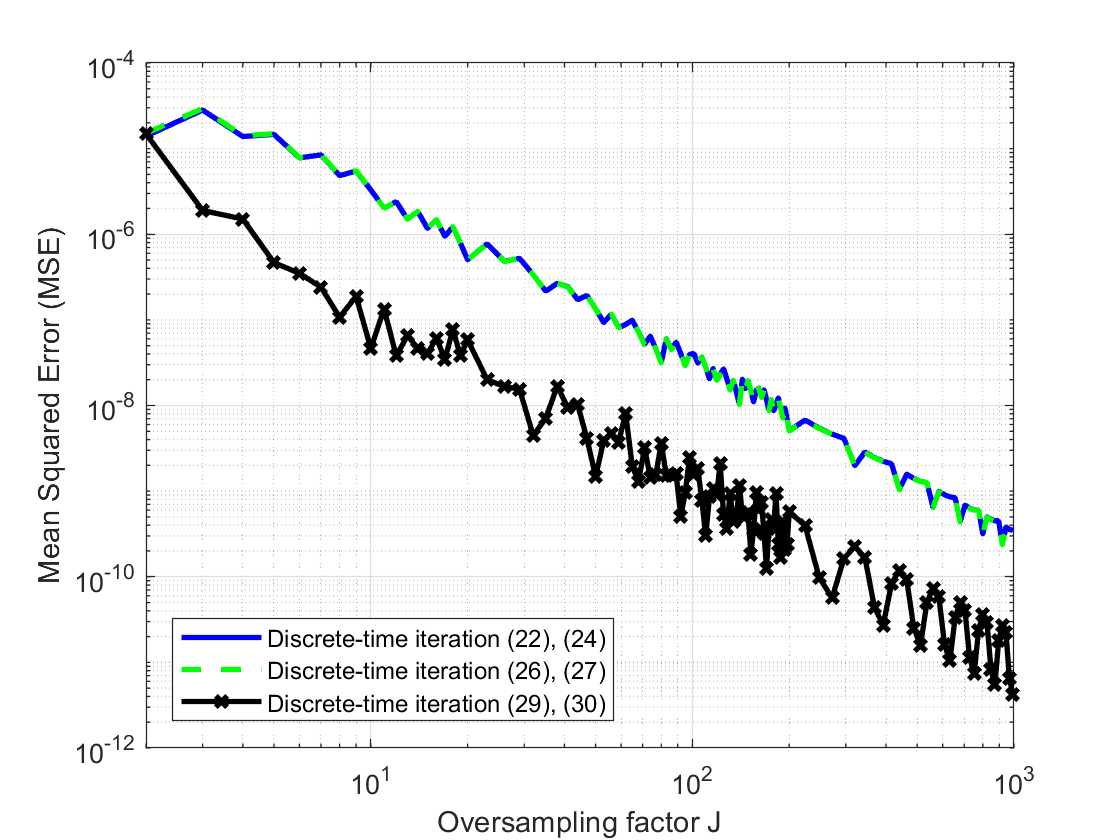}
\caption{Resulting MSE curves for the three discrete-time approximations as a function of the oversampling factor $J$ for the same setting as in Fig.~\ref{fig:v2_expl} and an example data sequence $x[k]=[1,0,1,0,1,...]$ of length 64.} 
\label{fig:mse}
\end{figure} 
The bias $\bias$ needs to be taken into account when evaluating the MSE results. To this end, the MSE is defined as 
\begin{equation}
{\sf MSE}:=\frac{1}{N_{\sf s}} \sum_{n=0}^{N_{\sf s}-1}\left(v_2[n]-\bias-v_{2,{\sf ref}}[n]\right)^2. 
\end{equation}
For the considered example, the resulting MSE curves are displayed in Fig.~\ref{fig:mse}. As can be seen,
above a certain waveform-dependent oversampling factor {(\mbox{$J>2$} in this example), the MSE is in essence inversely proportional to the oversampling factor $J$.  Hence, the error can be made arbitrarily small, but is never zero, because the voltage signal is not strictly bandlimited, cf.~Fig.~\ref{fig:V2_f_ripple}. Interestingly, while the MSE curves for the first two discrete-time iterations are virtually identical, the predictive model (\ref{eq:dot_iL_discrete_approx2024}), (\ref{eq:dot_v2_discrete_aprox2024}) exhibits a significantly better MSE curve in this example, although it contains further approximations compared to the first two discrete-time models. This means that the approximations $\alpha\approx 1$, $\beta\approx\Delta t/L$, $\gamma\approx-\beta$, $\kappa\approx(CR_{\sf L}-\Delta t)/(CR_{\sf L})$, and $\mu\approx\Delta t/C$ introduced in (\ref{eq:dot_iL_discrete_approx2021}), (\ref{eq:dot_v2_discrete_aprox2021}), which are also employed in the predictive model (\ref{eq:dot_iL_discrete_approx2024}), (\ref{eq:dot_v2_discrete_aprox2024}), have little impact on the resulting MSE curves in this case, while the additional approximations $s_1[n]\approx s_1[n-1]$ and $i_{\sf L}[n]\approx i_{\sf L}[n-1]$ employed in the predictive model obviously have a positive impact on the resulting MSE. The former finding is also confirmed by Fig.~\ref{fig:modelparam}, where the exact and the approximated values of the model parameters are displayed for a wide range of capacitance values $C$, corresponding inductance values\footnote{Note that the cut-off frequency $f_{\sf 3dB}$ according to (\ref{eq:f_3dB}) is identical for all considered cases, since the product $L\cdot C$ is kept constant.} of $L=1\cdot 10^{-11} {\rm \,H}\cdot{\rm F}/C$, an oversampling factor of $J=10$, and an example resistance value of $R_{\sf L}=20~\Omega$. Lines represent the exact values of the model parameters, and the markers represent the approximated values. As can be seen, the exact and approximated values are virtually identical, even though a relatively small oversampling factor $J$ has been chosen. Only for small capacitance values $C$, minor deviations are visible for parameter $\gamma$, as can be seen in the augmented subfigure.

Consequently, all discrete-time signal models have in common that the oversampling factor for modeling, $J$, must be sufficiently large, whereas for subsequent processing stages like equalization and detection small values like two or four are more appropriate. Therefore, $J$ should be a multiple of this when using the discrete-time model together with sub-sampling (see Section~\ref{VI-new-B} for further details).

\begin{figure}[t]
\centering
\includegraphics[width=0.5\textwidth]{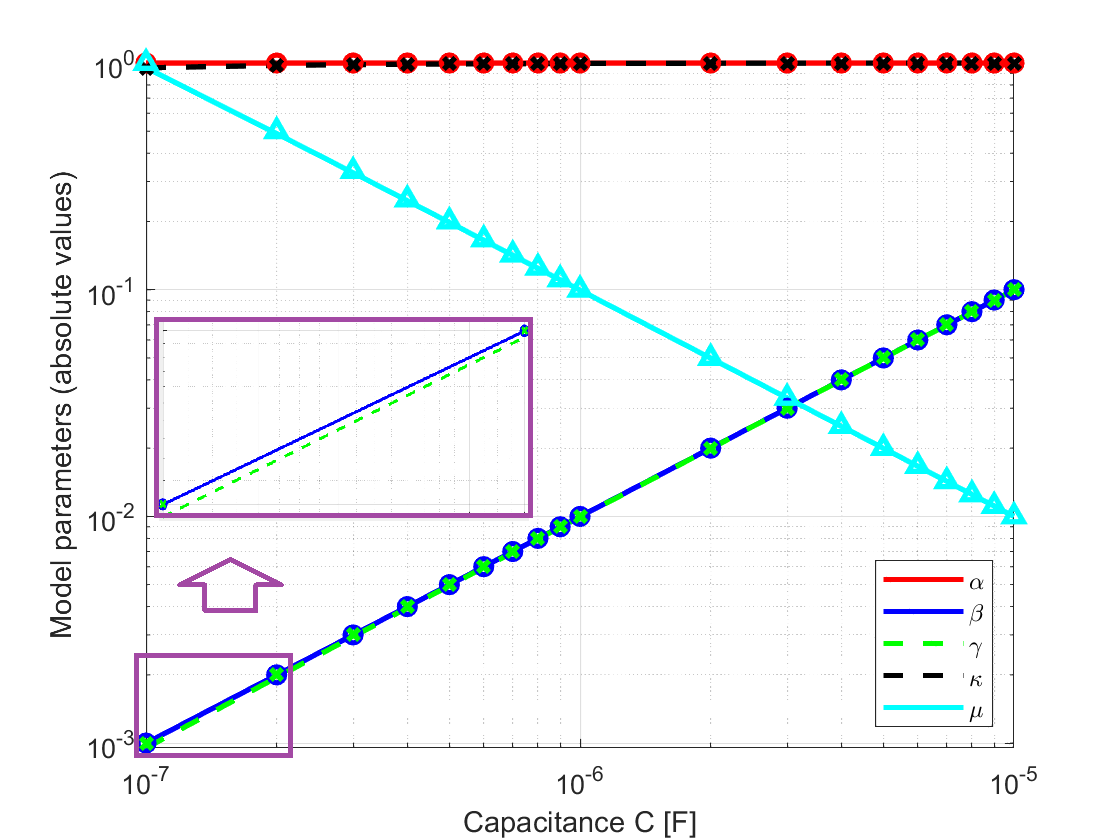}
\caption{Exact and approximated parameters of the discrete-time models for a wide range of capacitance values $C$, $LC=1\cdot 10^{-11} {\rm \,H}\cdot{\rm F}$, $J=10$, and $R_{\sf L}=20~\Omega$.} 
\label{fig:modelparam}
\end{figure} 

\subsection{Extension to Discontinuous Current Flow and Other Power Converters}
 A notable feature of discrete-time signal models is that an extension to discontinuous conduction mode (DCM) operation is straightforward. 
In DCM, there are time periods when the inductor current $i_{\sf L}(t)$ does not flow.  
DCM operation is only possible for asynchronous converters and for load resistances above a certain level. 
Compared to the synchronous DC/DC step-down converter in Fig.~\ref{fig:circuit_buck}, in the asynchronous variant the MOSFET switch $S_2$ is replaced by a diode. 
Correspondingly, the inductor current can only adopt positive values. 
Due to the lack of reverse recovery loss on the MOSFET's body diode, this potentially offers higher power conversion efficiency than for the CCM.
In order to model the DCM, the discrete-time models can be adapted by simply replacing $i_{\sf L}[n]$ with $[i_{\sf L}[n]]_+={\rm max}\{i_{\sf L}[n],0\}$. 
For example, in the iterative approximation (\ref{eq:dot_iL_discrete_acc}), (\ref{eq:dot_v2_discrete_acc}), the update rule (\ref{eq:dot_iL_discrete_acc}) translates to 
\begin{equation} 
    i_{\sf L}[n] = \bigg[\alpha \cdot i_{\sf L}[n-1] + \beta \cdot V_1\cdot s_1[n] + \gamma \cdot v_2[n-1]\bigg]_+   
\end{equation}
with initial values $i_{\sf L}[0]\geq 0$ and $v_2[0]\geq 0$. 
Note that the values for $v_2[n]$ will automatically remain positive in this case, as $\kappa,\mu>0$ holds.

Furthermore,  the general methodology of the proposed signal modeling concept is directly applicable to all types of switched-mode power
converters where the switching unit can be decoupled from the filter unit. If decoupling between switching and filtering is not possible, then an expanded system of equations needs to be set up. The main idea is to define state variables for all inductor currents and capacitor voltages. The resulting set of linear equations 
can then be solved recursively to update the state variables.

\section{Generic End-to-End Signal Model}\label{V}
Our analysis in Section~\ref{III}  was tailored to switching signals $s_1(t)$ with ideal rectangular shape. In particular, the overall pulse shape $g_{\sf tx}(t)$ in (\ref{eq:gtx_t}) contained both the influence of the modulation scheme and the circuit parameters. In practice, however, the switching signal $s_1(t)$ might exhibit a finite slope. It is therefore interesting to generalize the continuous-time model to arbitrary modulation signals $s_1(t)$. 

Since the input-/output behavior between $v_1(t)$ and $v_2(t)$ is linear and time-invariant, this is naturally done using the impulse response of the $LC$ lowpass filter with Ohmic load $R_{\sf L}$. The impulse response is the inverse Fourier transform of the frequency response $H(f)$ in (\ref{eq:Hf}) and is given by 
\begin{equation}\label{eq:hi}
    h_{\sf i}(t)=\frac{{\rm e}^{-at}\cdot {\rm sin}(bt)}{LCb}  \cdot \Theta(t). 
\end{equation}
As long as we can express $v_1(t)$ as a function of $s_1(t)$, we can directly obtain the data-dependent signal component of $v_2(t)$ through convolution with $h_{\sf i}(t)$. For example, it is easy to verify, e.g., by employing \cite[\S2.663, no.~1]{toi}, that the convolution of $v_1(t)=V_1\cdot s_1(t)$, with $s_1(t)$ according to (\ref{eq:s1}), and $h_{\sf i}(t)$ yields the expression (\ref{eq:v2d_t}) for $v_{2,{\sf d}}(t)$. 

The frequency response $H(f)$ and the impulse response $h_{\sf i}(t)$ -- or a sampled version thereof~-- may also serve as a basis for designing suitable analog and/or digital equalization techniques at the receiver, in order to effectively mitigate ISI effects and improve the resulting BER performance for TPC. Corresponding aspects are discussed in the following section. 

{\color{black}
\section{Implications of Continuous-Time and Discrete-Time Signal Modeling on Equalization
}\label{VI-new}

In the following, we provide a conceptional discussion, how our signal models can be utilized for developing corresponding receiver designs. Concrete BER examples are outside our scope, however, as they strongly depend on the targeted application and its specific preliminaries for the signal model (regarding, e.g., the choice of the PEC circuit parameters and the duty cycle), on the employed modulation and channel coding scheme, on the choice and the particular parametrization of the equalization scheme, and on additional receiver implementation issues (such as quantization, synchronization, and channel estimation). Corresponding detailed investigations are therefore delegated to future work.

\subsection{Implications of the Continuous-Time Signal Model}
The continuous-time signal model is very useful in order to predict the impact of the various involved parameters on the resulting signal behavior. This concerns the parameters of the DC-DC
step-down converter ($L$, $C$), the employed modulation scheme ($\delta$, $\tck$, $\Tpk$), as well as the Ohmic load $R_{\sf L}$. In particular, the final closed-form expression (\ref{eq:v2_final}) reveals that considerable ISI effects may occur. This can, for example, be seen when inspecting the resulting pulse shape $g_{\sf tx}(t)$, e.g., for the case \mbox{$L\!=\!10$~$\mu$H}, \mbox{$C\!=\!1$~$\mu$F}, \mbox{$R_{\sf L}\!=\!10$~$\Omega$}, which spans several symbol durations $T$ both for $\delta=0.5$ and $\delta=0.75$. This needs to be addressed at the receiver side. On the other hand, the ISI effect may be reduced considerably, by reducing the capacitance $C$ or the Ohmic load $R_{\sf L}$, cf.~Fig.~\ref{fig:gtx}. Note that the cut-off frequency of the $LC$ lowpass filter (and thus the available signal bandwidth) is identical for all cases considered in Fig.~\ref{fig:gtx}. 

At the receiver, ISI can be mitigated by employing a suitable equalization scheme. Known approaches will be based solely on the impulse response of the $LC$ lowpass filter with Ohmic load $R_{\sf L}$, see Section~\ref{V}, which disregards the transient behavior that occurs when switching, for example, from an unmodulated operation phase to a data transmission interval. Yet, the ripple behavior of the output voltage $v_2(t)$ is always present (even in the unmodulated case) \cite{Hoeher2021}, i.e., a steady-state situation ($v_2(0)=\delta\cdot V_1$, $\dot{v}_2(0) = 0$) at the start of a data transmission interval can typically not be assumed. 
Therefore, a receiver concept relying on standard equalization schemes will require transmitting a preamble prior to the start of the actual data transmission interval. This will allow the transient signal component $v_{\sf 2,t}(t)$ to taper out, 
but comes at the expense of a reduced data throughput. 

Our continuous-time signal model provides a blueprint for an alternative receiver concept: Prior to any equalization/ data detection steps, the transient component $v_{\sf 2,t}(t)$ may be subtracted at the receiver side. To this end, the initial values $v_2(0)$ and $\dot{v}_2(0)$ at the start of the data transmission interval need to be known, which leads to the central task of estimating $v_2(0)$ and $\dot{v}_2(0)$ at the receiver. In this context, our signal model can assist in analyzing the impact of possible estimation errors. Let 
\begin{eqnarray}
\hat{v}_2(0)&:=& v_2(0)+\varepsilon_{\sf v},\\
\hat{\dot{v}}_2(0) &:=& \dot{v}_2(0)+\varepsilon_{\sf dv}
\end{eqnarray}
denote the estimates of $v_2(0)$ and $\dot{v}_2(0)$, respectively, where $\varepsilon_{\sf v}$ and $\varepsilon_{\sf dv}$ denote corresponding error terms. The reconstructed transient signal component will thus read
\begin{eqnarray}
    \hat{v}_{\sf 2,t}(t)&=&{\rm e}^{-at}\bigg( \frac{ a\cdot \hat{v}_2(0) + \hat{\dot{v}}_2(0)}{b}\cdot {\rm sin}(bt)   \nonumber \\
    & &  \;\;\;\;\;\;\;\;\;\;\;\;+ \hat{v}_2(0)\cdot {\rm cos}(bt)\bigg)\cdot \Theta(t),
\end{eqnarray}
i.e., after subtraction the uncompensated transient signal component due to the estimation errors is given by 
\begin{equation}
\Delta v_{\sf 2,t}(t) = v_{\sf 2,t}(t) - \hat{v}_{\sf 2,t}(t)
\end{equation}
with $v_{\sf 2,t}(t)$ given by (\ref{eq:v2t_t}). This will allow to study the impact of the estimation errors $\varepsilon_{\sf v},\varepsilon_{\sf dv}$ on the resulting BER performance for any chosen equalization scheme.

{\color{black} 
Finally, we note that the frequency response $H(f)$ -- or, equivalently, the corresponding impulse response $h_{\sf i}(t)$ -- may directly be used for filter-based equalization schemes at the receiver, such as linear equalization or decision-feedback equalization \cite[Ch.~8]{BookFischerHuber}. For example, the frequency response of the zero-forcing linear equalizer is given by $F(f)=1/H(f).$ It may, in principle, be implemented either in the analog or in the digital domain~-- with respective approximations. In the digital case, frequency-domain processing is often the method of choice, via a corresponding fast-Fourier-transform operation.

\subsection{Implications of the Discrete-Time Signal Models}\label{VI-new-B}
The discrete-time signal models, on the other hand, are particularly useful for the design of digital state-based equalization schemes, such as maximum-likelihood or reduced-state sequence estimation techniques \cite{Hoeher2021},\cite[Ch.~8]{BookFischerHuber}. To this end, sub-sampled versions of the discrete-time signal models presented in Section~\ref{IV} may be employed, by reducing the oversampling factor $J$ by a factor of $N_{\sf sub}$, such that $S:=J/N_{\sf sub}$ is an integer value. In particular, $L_{\sf m}$ samples of the inductor current $i_{\sf L}(t)$ and the output voltage $v_2(t)$,
\[
\mathbf{i}_{\sf L}[m] := [i_{\sf L}[m],i_{\sf L}[m-1],\cdots,i_{\sf L}[m-L_{\sf m}+1]]
\]
\[
\mathbf{v}_2[m] := [v_2[m],v_2[m-1],\cdots,v_2[m-L_{\sf m}+1]],
\]
form a joint state ${\cal S}_m := [\mathbf{i}_{\sf L}[m],\mathbf{v}_2[m]]$ at time index $m$, where $i_{\sf L}[m]:=i_{\sf L}(mN_{\sf sub}\Delta t)$ and $v_2[m]:=v_2(mN_{\sf sub}\Delta t)$. Note that the state ${\cal S}_m$ is uniquely defined by the underlying data symbols $x[k]$, via (\ref{eq:dot_iL_discrete}) and (\ref{eq:dot_v2_discrete}). This facilitates corresponding sequence estimation at the receiver side, by comparing the observed samples of the output voltage $v_2(t)$ with (all) possible state sequences ${\cal S}_{m-\hat{D}},\cdots, {\cal S}_{m-1},{\cal S}_m$, where $\hat{D}$ denotes the decision delay.


Detailed investigations regarding suitable state-based equalization schemes for various use cases, such as block-based and continuous data transmission, are subject to future work. 
}

{\color{black}
\section{Generalization to Parasitic Effects and Arbitrary Impedance Loads}\label{VII-new}
To conclude the paper, we discuss two important generalization of the continuous-time signal model, which may be particularly relevant with regard to practical implementations: (i) parasitic effects and (ii) arbitrary impedance loads.

\subsection{Parasitic Effects}
Regarding parasitic effects, we may distinguish between the semiconductor switching devices and the passive circuit elements of the buck converter. Parasitic effects that influence the temporal behavior of the semiconductor switching devices can be
modeled by finite rise and fall times of the switching signal $s_1(t)$. To this end, the rectangular pulses in (\ref{eq:s1}) may be replaced by arbitrary pulses, e.g., pulses with exponential slopes. Utilizing the generic continuous-time signal model discussed in Section~\ref{V}, the data-dependent signal component $v_{2,{\sf d}}(t)$ of the output voltage $v_2(t)$ is readily obtained by convolving the new input signal $v_1(t)=V_1\cdot s_1(t)$ with the impulse response $h_{\sf i}(t)$ -- either analytically or numerically. Similarly, for the discrete-time signal models it is straightforward to substitute the rectangular pulses -- represented by the two-level samples $s_1[n]\in\{0,1\}$ -- by samples $s_1[n]$ representing an arbitrary pulse shape. MOSFET switching losses can be modeled by including additional resistances $R_{{\sf DS, ON},1}$ and $R_{{\sf DS, ON},2}$ in series with ideal switches $S_1$ and $S_2$, respectively.

Parasitic effects that concern the passive circuit elements of the buck converter can be taken into account by adding parasitic circuit elements. For the inductor $L$, three parasitic effects influencing its AC behavior are typically considered: an equivalent series resistance (ESR),
an equivalent parallel capacitance (EPC), and an equivalent parallel resistance (EPR). The main parasitic effects of the output capacitance $C$ are an ESR, an EPR, and an equivalent series inductor (ESL). In principle, all of these parasitic circuit elements may be included in Fig.~\ref{fig:circuit_buck}, and basic circuit analysis along the lines of Section~\ref{III} will then yield the corresponding DE, which may serve as a starting point for continuous-time or discrete-time signal modeling. In particular, the EPR of $C$ may be combined with the load resistance $R_{\sf L}$ to form a general impedance load~$X_{\sf L}$. 

An extended circuit model of the buck converter was considered in \cite{Chen2021}, where an ESL and an ESR was added to the output capacitance $C$. Yet, a comprehensive signal model for the output voltage $v_2(t)$ -- including the transient signal behavior~-- was not presented. 
It can be shown that the extension proposed in \cite{Chen2021} leads to a new expression for $V_2(s)$, see (\ref{eq:V2s_extcase1}), where
\[
K(s) :=K_2 s^2 + K_1 s + K_0
\]
\[
M_{0}(s) := M_{0,1}s+M_{0,0}
\]
\[
N_{0}(s) := N_{0,2}s^2+N_{0,1}s+N_{0,0}
\]
\[
N_{1}(s) := N_{1,1}s+N_{1,0}
\]
denote polynomials of order $\{2,1,2,1\}$, respectively, with coefficients depending on the choice of $L,C,R_{\sf L}$, the ESR and the ESL. Compared to (\ref{eq:Laplace_V2}) we note that there is an additional polynomial $K(s)$ of order two in the numerator of $V_2(s)$, which is associated with $V_1(s)$. Furthermore, the initial conditions $v_1(0)$ and $\dot{v}_1(0)$ of the voltage at the input of the filter now come into play. Finally, we note that the denominator polynomial of $V_2(s)$ has increased from order two to order three.

\begin{figure*}[t!] 
\normalsize
\setcounter{mytempeqncnt}{\value{equation}}
\setcounter{equation}{38}
\begin{eqnarray}\label{eq:V2s_extcase1}
 V_2(s)&=&\frac{K(s)\cdot V_1(s) + M_{0}(s)\cdot v_1(0) + M_{1,0}\cdot \dot{v}_1(0)+ N_{0}(s)\cdot v_2(0) + N_{1}(s)\cdot \dot{v}_2(0) + N_{2,0}\cdot \ddot{v}_2(0)}{(s-s_{0,1}) (s-s_{0,2})(s-s_{0,3})}
\end{eqnarray}
\setcounter{equation}{39}
\hrulefill
\vspace*{4pt}
\end{figure*} 

\begin{figure*}[t!] 
\normalsize
\setcounter{mytempeqncnt}{\value{equation}}
\setcounter{equation}{39}
\begin{eqnarray}\label{eq:V2s_extcase2}
 V_2(s)&=&\frac{\tilde{K}(s)\cdot V_1(s) + \tilde{M}_{0}(s)\cdot v_1(0) + \tilde{M}_{1,0}\cdot \dot{v}_1(0)+ \tilde{N}_{0}(s)\cdot v_2(0) + \tilde{N}_{1}(s)\cdot \dot{v}_2(0) + \tilde{N}_{2}(s)\cdot \ddot{v}_2(0) + \tilde{N}_{3,0}\dddot{v}_2(0)}{(s-s_{0,1}) (s-s_{0,2})(s-s_{0,3})(s-s_{0,4})}\nonumber \\
\end{eqnarray}
\setcounter{equation}{40}
\hrulefill
\vspace*{4pt}
\end{figure*} 

\subsection{Arbitrary Impedance Loads}
Concerning arbitrary impedance loads, we consider another extension of the basic circuit model in Fig.~\ref{fig:circuit_buck} in more detail, namely an extension to an arbitrary impedance load $X_{\sf L}$ composed of an Ohmic load $R_{\sf L}$, a series inductance load $L_{\sf L}$, and a series capacitance load $C_{\sf L}$. For the ease of exposition, we assume ideal conditions regarding the elements $L$ and $C$ of the buck converter circuit, i.e., parasitic effects are disregarded. It can be shown that this generalization leads to a new expression for $V_2(s)$, see (\ref{eq:V2s_extcase2}), where
\[
\tilde{K}(s) :=\tilde{K}_2 s^2 + \tilde{K}_1 s + \tilde{K}_0
\]
\[
\tilde{M}_{0}(s) := \tilde{M}_{0,1}s+\tilde{M}_{0,0}
\]
\[
\tilde{N}_{0}(s) := \tilde{N}_{0,3}s^3+\tilde{N}_{0,2}s^2+\tilde{N}_{0,1}s+\tilde{N}_{0,0}
\]
\[
\tilde{N}_{1}(s) := \tilde{N}_{1,2}s^2+\tilde{N}_{1,1}s+\tilde{N}_{1,0}
\]
\[
\tilde{N}_{2}(s) := \tilde{N}_{2,1}s+\tilde{N}_{2,0}
\]
denote polynomials of order $\{2,1,3,2,1\}$, respectively, with coefficients depending on the choice of $L,C,R_{\sf L},C_{\sf L}$, and $L_{\sf L}$. Compared to (\ref{eq:V2s_extcase1}) we note that the orders of the polynomials in the numerator of $V_2(s)$, which are associated with the initial conditions of the output voltage $v_2(t)$, have increased from a maximum of two to a maximum of three. Furthermore, the degree of the denominator polynomial has increased from three to four.
An analytical solution for the output voltage $v_2(t)$ may be obtained along the lines of Section~\ref{III}, by rewriting (\ref{eq:V2s_extcase2}) in terms of a partial fraction expansion 
\begin{equation}
    V_2(s) = \frac{\tilde{A}}{s-s_{0,1}} + \frac{\tilde{B}}{s-s_{0,2}} + \frac{\tilde{C}}{s-s_{0,3}} + \frac{\tilde{D}}{s-s_{0,4}},
\end{equation}
where we have assumed that the resulting zeros $s_{0,1}, ..., s_{0,4}$ of the denominator polynomial (i.e., the poles of $V_2(s)$) are distinct. Note that there are analytical methods to find the zeros of a general polynomial up to a degree of four \cite{Chavez-Pichardo2023}. The coefficients $\tilde{A}$, $\tilde{B}$, $\tilde{C}$, $\tilde{D}$  
 need to be evaluated based on the given polynomials $\tilde{K}(s)$, $\tilde{M}_{0}(s)$, and $\tilde{N}_{i}(s)$ ($i=0,1,2$), the poles $s_{0,1}, ..., s_{0,4}$ of $V_2(s)$, and the initial conditions for $v_1(t)$ and $v_2(t)$.  $V_2(s)$ may finally be transformed back to the time domain, by using standard Laplace transform pairs.


}

\section{Conclusion and Outlook}\label{VI}
We devised a comprehensive signal model for joint information and power transfer based on the concept of TPC. 
Focus was on the example of a synchronous step-down DC/DC voltage converter with an $LC$ lowpass filter and an Ohmic load. 
The output voltage of this buck converter was shown to consist of a transient and a data-dependent component, and we analyzed the resulting information-carrying voltage ripple for a broad range of pulsed-based modulation schemes. 
In particular, the influence of the modulation and circuit parameters was studied in both time and frequency domains.  

Starting from the continuous-time signal model, we derived several discrete-time signal models and assessed their respective accuracies. 
It was shown that discrete-time signal models can introduce a bias regarding the average output voltage. 
Furthermore, the (bias-corrected) MSE compared to the continuous-time signal model can be made arbitrarily small by increasing the oversampling factor. 
A notable advantage of the discrete-time signal models is that they can readily be used to simulate TPC under DCM operation of asynchronous DC/DC voltage converters. 
Finally, we discussed a generic end-to-end continuous-time signal model, which is valid for arbitrary modulation signals and may serve as a basis for designing suitable analog and/or digital equalization techniques, in order to mitigate ISI effects at the receiver side. In order to provide further insight, we highlighted the role of the transient-signal component for the receiver design and discussed implications of our proposed
continuous-time and discrete-time signal modeling on equalization. We furthermore generalized our continuous-time signal modeling approach to include parasitic effects as well as arbitrary impedance loads.

Although our focus has been on the buck converter throughout this article, the general methodology of the proposed signal modeling concept is applicable to all types of switched-mode power converters where the switching unit can be decoupled from the filter unit. Apart from the buck converter this applies, for example, to the phase-shifted full-bridge converter. Counter-examples include the boost converter and the flyback converter \cite{BookKazimierczuk}.

Interesting directions for future work include more detailed investigations regarding general impedance loads as well as the impact of parasitic effects and non-ideal switches with finite switching times for various use cases. Additionally, exploring alternative analog lowpass filters for the PEC may be of interest.

As long as a DE (which accurately models the overall behavior) can be formulated, it will always be feasible to derive a corresponding discrete-time signal model~-- possibly, with a more complicated structure. If parts of the overall circuit model are not precisely known, however, it might be favorable to avoid rigorous mathematical signal modeling and rather apply, for example, suitable machine-learning techniques, in order to learn the overall system behavior and adjust transmitter- and/or receiver-sided signal processing accordingly. Such considerations will be subject of future work.

\section*{Acknowledgment}
C. Chakraborty and L. Lampe acknowledge support by the Natural Sciences and Engineering Research Council (NSERC) of Canada. 

{\color{black}
\section*{Appendix}
In this appendix, we provide the details for the derivation of Equations (\ref{eq:V1_s_expl}) to (\ref{eq:gtx_t}) presented in Section~\ref{III}.

Using the definitions for the start $\tsk$ and the end $\tek$ of the $k$th pulse, and constraining $s_1(t)$ to a right-sided signal over $K$ symbol periods ($0\leq k\leq K-1$), we may write
\begin{equation}
    s_1(t) = \sum_{k=0}^{K-1} \Big( \Theta(t-\tsk-kT) - \Theta(t-\tek-kT) \Big).
\end{equation}
The Laplace-transform of $\Theta(t-t_0)$ is given by ${\rm e}^{-st_0}/s$. Employing $v_1(t) = V_1\cdot s(t)$, we thus arrive at the Laplace transform (\ref{eq:V1_s_expl}) of the voltage $v_1(t)$ delivered by the PEC switching unit. Plugging (\ref{eq:V1_s_expl}) into (\ref{eq:Laplace_V2}) yields the corresponding expression (\ref{eq:V2_s_expl}) for the Laplace transform of the output voltage $v_2(t)$ of the step-down converter. Here, $s_{0,1/2}$ are the distinct and complex-conjugated poles of $V_2(s)$, which are of form $s_{0,1}=-a+{\rm j}b$ and $s_{0,2}=-a-{\rm j}b$.

The first term in (\ref{eq:V2_s_expl}), denoted as $V_{2,{\sf t}}(s)$ in the sequel, concerns the transient
behavior of the output voltage $v_2(t)$ starting from the initial conditions $v_2(0)$ and $\dot{v}_2(0)$. It may be rewritten as
\begin{equation}
V_{2,{\sf t}}(s) = \frac{A}{s-s_{0,1}} + \frac{B}{s-s_{0,2}},  
\end{equation}
by means of a partial fraction expansion, where
\begin{equation}
A = \frac{v_2(0)\cdot (s_{0,1}+1/(R_{\sf L}C)) +\dot{v}_2(0)}{s_{0,1}-s_{0,2}}  
\end{equation}
and
\begin{equation}
B = -\frac{v_2(0)\cdot (s_{0,2}+1/(R_{\sf L}C)) +\dot{v}_2(0)}{s_{0,1}-s_{0,2}} . 
\end{equation}
Noting that $1/(s-\alpha)$ is the Laplace transform of ${\rm e}^{\alpha t}\cdot \Theta(t)$, with convergence for any $s\in \mathbb{C}$ for which ${\rm Re}\{s+\alpha\}>0$ holds ($\alpha\in \mathbb{C}$), we get
\begin{equation}
v_{2,{\sf t}}(t) = \left(A\cdot{\rm e}^{s_{0,1} t} + B\cdot{\rm e}^{s_{0,2} t}\right)\cdot \Theta(t)  
\end{equation}
for the corresponding time-domain signal. Inserting $s_{0,1,2}=-a\pm{\rm j}b$ and invoking Euler's identities ${\rm cos}(x)= ({\rm e}^{{\rm j}x}+{\rm e}^{-{\rm j}x})/2$ and ${\rm sin}(x)= ({\rm e}^{{\rm j}x}-{\rm e}^{-{\rm j}x})/(2{\rm j})$, we finally arrive at (\ref{eq:v2t_t}).

The second term in (\ref{eq:V2_s_expl}) concerns the data-dependent
behavior of the output voltage $v_2(t)$. Using that a product in the $s$-domain corresponds to a linear convolution in time-domain, and invoking the Laplace transform pairs used above, we get
\begin{eqnarray}
v_{2,{\sf d}}(t) &\!\!\!=\!\!\!& \frac{V_1}{LC}\cdot\left( \Theta(t) \ast {\rm e}^{s_{0,1} t}\cdot \Theta(t) \ast {\rm e}^{s_{0,2} t}\cdot \Theta(t)  \right)      \\ 
&\!\!\!\cdots\!\!\!  & \ast \sum_{k=0}^{K-1} \delta_0(t-\tsk-kT) - \delta_0(t-\tek-kT) \nonumber
\end{eqnarray}
for the corresponding time-domain signal. The second convolution may be expressed as 
\begin{equation}
{\rm e}^{s_{0,1}t}\cdot \Theta(t) \ast {\rm e}^{s_{0,2}t}\cdot \Theta(t) = \frac{{\rm e}^{s_{0,1}t}-{\rm e}^{s_{0,2}t}}{s_{0,1}-s_{0,2}} \cdot \Theta(t). 
\end{equation} 
By solving the convolution integral of form 
\begin{eqnarray}
\Theta(t)\ast {\rm e}^{\alpha t}\cdot \Theta(t)&=&\int_{-\infty}^{+\infty} \Theta(\tau)\cdot {\rm e}^{\alpha (t-\tau)}\cdot \Theta(t-\tau){\rm d}\tau \nonumber \\
& = &\left( {\rm e}^{\alpha t} - 1 \right)/\alpha \cdot \Theta(t) , 
\end{eqnarray} 
we obtain
\begin{eqnarray}
v_{2,{\sf d}}(t) &\!\!\!=\!\!\!& \frac{V_1}{LC(s_{0,1}-s_{0,2})}\left( \frac{{\rm e}^{s_{0,1} t}-1}{s_{0,1}}-\frac{{\rm e}^{s_{0,2} t}-1}{s_{0,2}}  \right)\cdot \Theta(t)  \nonumber \\ 
\!\!\!&\!\!\!\!\!\!\cdots\!\!\!  & \ast \sum_{k=0}^{K-1} \delta_0(t-\tsk-kT) - \delta_0(t-\tek-kT). \nonumber \\
\end{eqnarray}
Inserting $s_{0,1,2}=-a\pm{\rm j}b$, invoking Euler's identities for ${\rm cos}(x)$ and ${\rm sin}(x)$, and using the identities 
\[
s_{0,1}\cdot s_{0,2}=a^2+b^2=1/(LC),
\]
we finally arrive at the closed-form expression (\ref{eq:v2d_t}) for $v_{2,{\sf d}}(t)$.

Furthermore, by exchanging the convolution operation and the summation in (\ref{eq:v2d_t}), we may write
\begin{eqnarray} 
    &&\hspace*{-0.5cm} v_{2,{\sf d}}(t)= \\
    &&\hspace*{0.1cm} V_1\cdot \sum_{k=0}^{K-1}\left( 1-{\rm e}^{-at} \cdot \left({\rm cos}(bt) + \frac{a}{b}\cdot {\rm sin}(bt) \right)\right)\cdot \Theta(t)  \nonumber \\
    &&\hspace*{0.5cm} \cdots \ast \Big(   \delta_0(t-kT-\tsk) - \delta_0(t-kT-\tek)\Big),\nonumber
\end{eqnarray}
and by using the identity $x(t)\ast \delta_0(t-t_0) = x(t-t_0)$, which holds for any arbitrary signal $x(t)$, we arrive at (\ref{eq:v2d_pulses}) with basic pulse shape $g_{\sf tx}(t)$ as given in (\ref{eq:gtx_t}).

}


\begin{IEEEbiography}[{\includegraphics[width=1in,height=1.25in,clip,keepaspectratio]{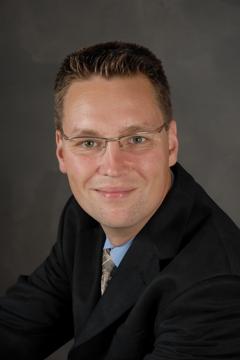}}]{Jan Mietzner\,} (Senior Member, IEEE) received his Dr.-Ing.~degree (Hons.) in electrical and information engineering from Kiel University, Germany, in 2006. From 2007 to 2008, he was a Postdoctoral Research Fellow with The University of British Columbia, Vancouver, BC, Canada, sponsored by the German Academic Exchange Service (DAAD), and returned in 2023 as a Visiting Professor for a research semester. In 2009, he joined Airbus DS (now Hensoldt), Ulm, Germany, where he worked in the areas of jamming and radar systems. In September 2017, he became a Professor of communications engineering with the Hamburg University of Applied Sciences (HAW). His research interests include theoretical and practical aspects of communication and radar systems. He has published more than 30 journal articles and 50 conference papers. He was a co-recipient of the 2010 Best Paper Award from the German Information Technology Society (VDE/ITG). He has served as a Technical Program Committee Member for various IEEE conferences and was the Track Co-Chair of VTC-Fall 2014. He also served as an Editor for the IEEE Wireless Communications Letters.\\[-4ex]
\end{IEEEbiography}

\begin{IEEEbiography}[{\includegraphics[width=1in,height=1.25in,clip,keepaspectratio]{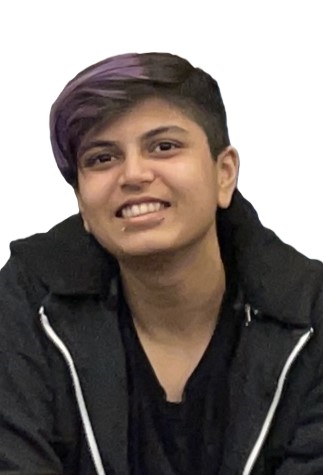}}]{Cerikh Chakraborty\,}(Graduate Student Member, IEEE) received the integrated B.Tech. and M.Tech. dual degree from Indian Institute of Technology Kharagpur, Kharagpur, India, in 2023 and is currently pursuing the MASc. degree at the Department of Electrical and Computer Engineering, The University of British Columbia, Vancouver, BC, Canada. Their research interests include talkative power conversion (TPC) and corresponding receiver schemes, as well as machine-learning techniques for TPC.\\[-4ex]  
\end{IEEEbiography}

\begin{IEEEbiography}[{\includegraphics[width=1in,height=1.25in,clip,keepaspectratio]{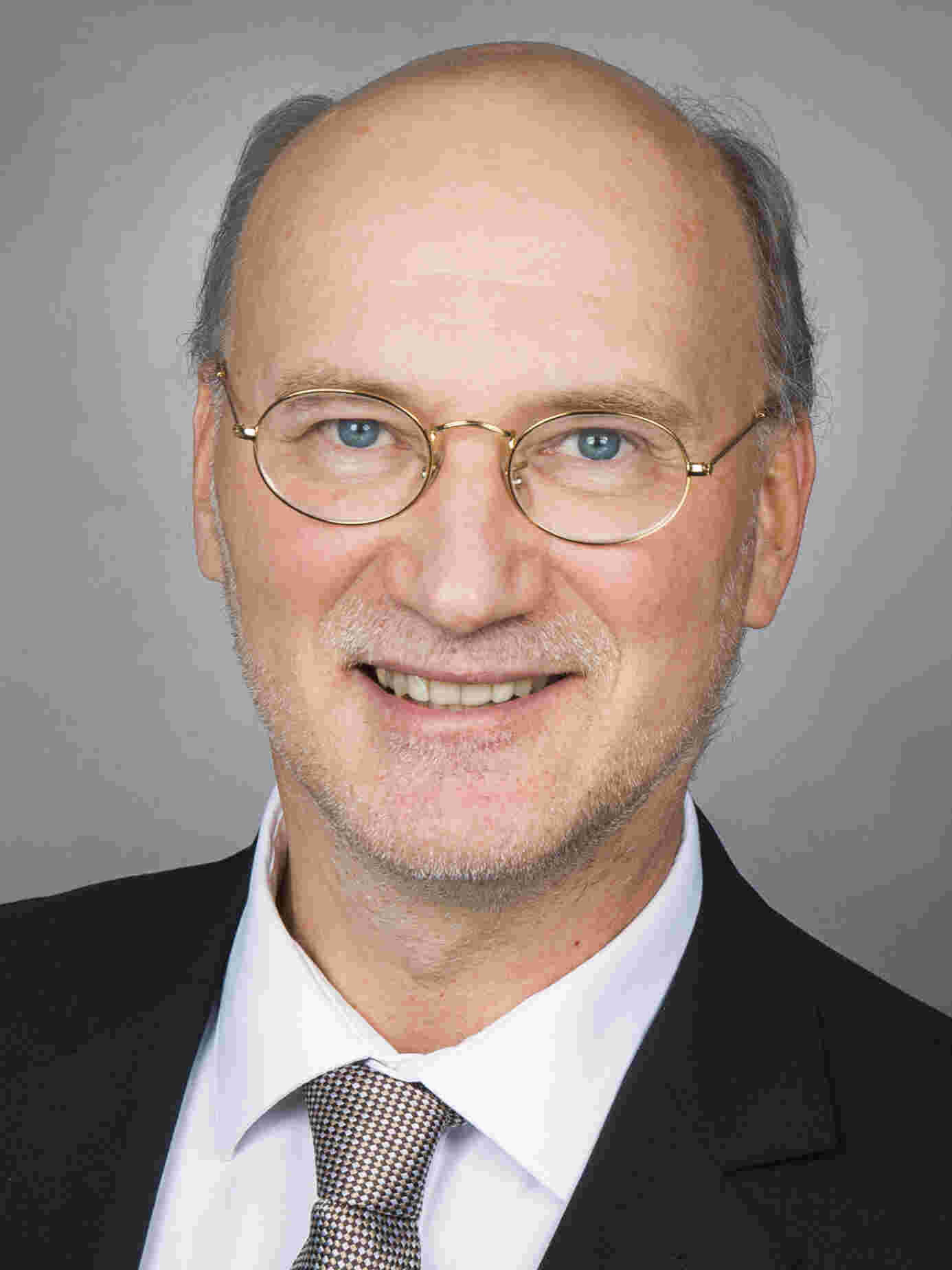}}]{Peter A. Hoeher\,}(Fellow, IEEE) received the Dipl.-Ing. degree in electrical engineering from RWTH Aachen University, Aachen, Germany, in 1986, and the Dr.-Ing. degree in electrical engineering from the University of Kaiserslautern, Kaiserslautern, Germany, in 1990.
From 1986 to 1998, he was with the German Aerospace Center (DLR), Oberpfaffenhofen, Germany. From 1991 to 1992, he was on leave at AT\&T Bell Laboratories, Murray Hill, NJ, USA. In 1998, he joined Kiel University, Kiel, Germany, where he is currently a Full Professor of electrical and information engineering. His research interests are in the general area of communication theory and applied information theory with applications in wireless radio communications, optical wireless communications, and advanced communications in power electronics.
Since 2014, Dr. Hoeher has been an IEEE Fellow for contributions to decoding and detection that include reliability information. From 1999 to 2006, he served as an Associated Editor for the IEEE Transactions on Communications.\\[-4ex]
\end{IEEEbiography}

\begin{IEEEbiography}[{\includegraphics[width=1in,height=1.25in,clip,keepaspectratio]{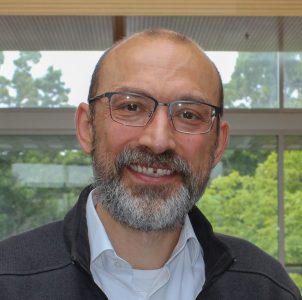}}]{Lutz Lampe\,}(Senior Member, IEEE) received the Dipl.-Ing.\ and Dr.-Ing.\ degrees in electrical engineering from the University of Erlangen, Erlangen, Germany, in 1998 and 2002, respectively. Since 2003, he has been with the Department of Electrical and Computer Engineering, The University of British Columbia, Vancouver, BC, Canada, where he is a Full Professor. He is a Co-Editor of the book “Power Line Communications: Principles, Standards and Applications from Multimedia to Smart Grid” (2nd edition, John Wiley \& Sons). His research interests are broadly in theory and application of wireless, optical wireless, optical fiber, power line, and underwater acoustic communications. He has been a (co-)recipient of a number of best paper awards. He has served as an associate editor and a guest editor for
several IEEE journals, and as the general and technical program committee co-chair for IEEE conferences. He has been a Distinguished Lecturer of the IEEE Communications Society.
\end{IEEEbiography}
\vfill
\end{document}